\documentclass[twocolumn]{aastex62}

\newcommand{\eg}{e.g., }
\newcommand{\ie}{i.e., }
\newcommand{\Msun}{M_{\odot}}
\newcommand{\Rsun}{R_{\odot}}

\newcommand{\Rs}{R_{\rm preSN}}

\newcommand{\Mms}{M_{\rm ZAMS}}

\newcommand{\lres}{{\lambda_{\rm rest}}}
\newcommand{\SNa}{{SHOOT14di}}
\newcommand{\RADECa}{${\rm R.A.=21^{h}33^m04.\hspace{-.55ex}^s27}$, ${\rm decl.=+09^\circ35'55.\hspace{-.5ex}''0}$ (J2000.0)}
\newcommand{\RADEChost}{${\rm R.A.=21^{h}33^m04.\hspace{-.55ex}^s30}$, ${\rm decl.=+09^\circ35'54.\hspace{-.5ex}''9}$ (J2000.0)}

\newcommand{\SNsnls}{SNLS-04D2dc}
\newcommand{\SNps}{PS1-13arp}
\newcommand{\SNkepler}{KSN~2011d}
\newcommand{\SNmelina}{SN~2016gkg}

\newcommand{\Ebvg}{E_{B-V, {\rm Gal}}}
\def\gsim{\mathrel{\rlap{\lower 4pt \hbox{\hskip 1pt $\sim$}}\raise 1pt
\hbox {$>$}}}
\def\lsim{\mathrel{\rlap{\lower 4pt \hbox{\hskip 1pt $\sim$}}\raise 1pt
\hbox {$<$}}}

\newcommand{\Rd}{R({{\rm 0.7day})}}
\newcommand{\Rdps}{R({2000{\rm \AA},{\rm 0.7day})}}
\newcommand{\Rgd}{R({3400{\rm \AA},{\rm 0.70day})}}

\newcommand{\Junit}{\mu{\rm Jy}}
\newcommand{\Runit}{\mu{\rm Jy~day^{-1}}}
\newcommand{\Munit}{{\rm mag~day^{-1}}}
\newcommand{\Mg}{M_{\rm 3400\AA}}
\newcommand{\Mr}{M_{\rm 4400\AA}}

\received{June 16, 2019}
\accepted{September 5, 2019}

\begin{document}

\title{A rapidly declining transient discovered with Subaru/Hyper Suprime-Cam}

\author{Nozomu~Tominaga}
\affiliation{Department of Physics, Faculty of Science and
Engineering, Konan University, 8-9-1 Okamoto,
Kobe, Hyogo 658-8501, Japan; tominaga@konan-u.ac.jp}
\affiliation{Kavli Institute for the Physics and Mathematics of the
Universe (WPI), The University of Tokyo, 5-1-5 Kashiwanoha, Kashiwa, Chiba
277-8583, Japan}

\author{Tomoki~Morokuma}
\affiliation{Institute of Astronomy, Graduate School of Science, The
University of Tokyo, 2-21-1 Osawa, Mitaka, Tokyo 181-0015, Japan}
\affiliation{Kavli Institute for the Physics and Mathematics of the
Universe (WPI), The University of Tokyo, 5-1-5 Kashiwanoha, Kashiwa, Chiba
277-8583, Japan}

\author{Masaomi~Tanaka}
\affiliation{Astronomical Institute, Tohoku University, Aoba, Sendai, Miyagi 980-8578, Japan}
\affiliation{Kavli Institute for the Physics and Mathematics of the
Universe (WPI), The University of Tokyo, 5-1-5 Kashiwanoha, Kashiwa, Chiba
277-8583, Japan}

\author{Naoki~Yasuda}
\affiliation{Kavli Institute for the Physics and Mathematics of the
Universe (WPI), The University of Tokyo, 5-1-5 Kashiwanoha, Kashiwa, Chiba
277-8583, Japan}

\author{Hisanori~Furusawa}
\affiliation{National Astronomical Observatory of Japan, 2-21-1
Osawa, Mitaka, Tokyo 181-8588, Japan}

\author{Masayuki~Tanaka}
\affiliation{National Astronomical Observatory of Japan, 2-21-1
Osawa, Mitaka, Tokyo 181-8588, Japan}

\author{Ji-an~Jiang}
\affiliation{Kavli Institute for the Physics and Mathematics of the
Universe (WPI), The University of Tokyo, 5-1-5 Kashiwanoha, Kashiwa, Chiba
277-8583, Japan}
\affiliation{Institute of Astronomy, Graduate School of Science, The
University of Tokyo, 2-21-1 Osawa, Mitaka, Tokyo 181-0015, Japan}

\author{Alexey~Tolstov}
\affiliation{The Open University of Japan, 2-11, Wakaba, Mihama-ku, Chiba, Chiba
261-8586, Japan}

\author{Sergei~Blinnikov}
\affiliation{NRC ``Kurchatov institute'' - ITEP, Moscow 117218, Russia}
\affiliation{Kavli Institute for the Physics and Mathematics of the
Universe (WPI), The University of Tokyo, 5-1-5 Kashiwanoha, Kashiwa, Chiba
277-8583, Japan}
\affiliation{SAI MSU Moscow 119234, Russia}

\author{Mamoru~Doi}
\affiliation{Institute of Astronomy, Graduate School of Science, The
University of Tokyo, 2-21-1 Osawa, Mitaka, Tokyo 181-0015, Japan}
\affiliation{Research Center for the Early Universe, Graduate School of Science, The 
University of Tokyo, 7-3-1, Hongo, Bunkyo-ku, Tokyo 113-003, Japan}

\author{Ikuru~Iwata}
\affiliation{The Graduate University for Advanced Studies (SOKENDAI),
Tokyo 181-8588, Japan}
\affiliation{Subaru Telescope, National Astronomical
Observatory of Japan, 650 North A'ohoku Place, Hilo, HI 96720, USA}
\affiliation{Department of Astronomy and Physics and Institute for Computational
Astrophysics, Saint Mary's University, 923 Robie Street, Halifax, Nova
Scotia B3H 3C 3, Canada}

\author{Hanindyo~Kuncarayakti}
\affiliation{Finnish Centre for Astronomy with ESO (FINCA), FI-20014
University of Turku, Finland}
\affiliation{Tuorla Observatory, Department of Physics and Astronomy,
FI-20014 University of Turku, Finland}

\author{Takashi~J.~Moriya}
\affiliation{National Astronomical Observatory of Japan, 2-21-1
Osawa, Mitaka, Tokyo 181-8588, Japan}

\author{Tohru~Nagao}
\affiliation{Research Center for Space and Cosmic Evolution, Ehime
University, Bunkyo-cho, Matsuyama 790-8577, Japan}

\author{Ken'ichi~Nomoto}
\affiliation{Kavli Institute for the Physics and Mathematics of the
Universe (WPI), The University of Tokyo, 5-1-5 Kashiwanoha, Kashiwa, Chiba
277-8583, Japan}
\altaffiliation{Hamamatsu Professor}

\author{Junichi~Noumaru}
\affiliation{Subaru Telescope, National Astronomical
Observatory of Japan, 650 North A'ohoku Place, Hilo, HI 96720, USA}

\author{Tadafumi~Takata}
\affiliation{National Astronomical Observatory of Japan, 2-21-1
Osawa, Mitaka, Tokyo 181-8588, Japan}

 \begin{abstract}
 We perform a high-cadence transient survey with Subaru Hyper
 Suprime-Cam (HSC), 
which we call the Subaru HSC survey Optimized for Optical Transients
 (SHOOT). We conduct HSC imaging observations with time intervals of
 about one hour on two successive nights, and spectroscopic and photometric
 follow-up observations. 
 A rapidly declining blue transient \SNa\ at $z=0.4229$ is found in 
 observations on two successive nights with an image subtraction technique. 
 The rate of brightness
  change is $+1.28^{+0.40}_{-0.27}~\Munit$
  ($+1.83^{+0.57}_{-0.39}~\Munit$) in the observer (rest) frame
  and the 
    rest-frame 
  color between $3400$ and $4400$~\AA\ is
  $\Mg-\Mr=-0.4$. The nature of the object is
 investigated by comparing its peak luminosity, decline rate, and color 
 with those of transients and variables previously observed,
  and those of theoretical models. None
  of the transients or variables share the same properties as
  \SNa. Comparisons with theoretical models demonstrate that, while 
     the emission from the cooling
  envelope of a Type IIb supernova shows a slower decline rate than
  \SNa, and the explosion of a red supergiant star with a dense
  circumstellar wind shows a redder color than \SNa, the shock breakout at
  the stellar surface of the explosion of a $25\Msun$ red supergiant 
  star with 
  a small
  explosion energy of $\leq0.4\times10^{51}$~erg reproduces the
  multicolor light curve of \SNa. This discovery shows that a high-cadence,
  multicolor optical transient survey at intervals of about one hour,
  and continuous and immediate follow-up observations, is important for studies of
  normal core-collapse supernovae at high redshifts.
 \end{abstract}

\keywords{shock waves --- radiative transfer --- supernovae: general ---
supernovae: individual (\SNa) --- stars: evolution --- surveys}

\section{INTRODUCTION}
\label{sec:intro}

Traditional transient surveys have been performed mainly with
a cadence of several days in order to catch large numbers of Type Ia
supernovae. 
Searching the transient sky at a shorter timescale 
   has gained attention as a new
frontier of astronomy in this decade. In particular, time scales as short as 
$1$~day have been intensively investigated by transient surveys with wide-field
cameras, \eg the Palomar Transient Factory (PTF, \citealt{law09,rau09})
and Zwicky Transient Facility (ZTF, \citealt{gra19ZTF}) with a $1.2$~m
telescope, the Catalina Real-Time Transient Survey with a $0.7$~m telescope
(CRTS, \citealt{dra09}), the Kiso Supernova Survey with a $1$~m telescope
(KISS, \citealt{mor14}), and the High-cadence Transient Survey with a $4$~m
telescope (HiTS, \citealt{for16}).

Several phenomena are theoretically proposed to appear
in ultraviolet and optical
bands at the short time scale of $\lsim1$~day and will offer new
insights, especially on the final stages of the evolution of massive
stars. For example, the emission from a shock breakout at the stellar
surface \citep{kle78} and in the dense wind \citep{che11} and 
subsequent emission from the cooling envelopes \citep{wax07,nak10} of 
core-collapse supernovae of red supergiant stars, have characteristic
time scales of $1$~hr to several days. 
Studying such transients will reveal the stellar
radius and the structure of the circumstellar medium surrounding the
progenitor star, and thus the mass loss just before the core
collapse.

Ultraviolet observations with the {\it GALEX} satellite
\citep{mor05,mor07} reveal a brightening at the position of Type II
plateau supernovae; \SNsnls\ ($z=0.185$,
\citealt{sch08,gez08}), SNLS-06D1jd ($z=0.324$, \citealt{gez08}), and
\SNps\ ($z=0.1665$, \citealt{gez15}). The UV brightening of
\SNsnls\ is well reproduced by the emission of the shock breakout at the
stellar surface of a star with a zero-age main sequence mass $\Mms$ of $20\Msun$,
solar metallicity, \ie $\Rs=800\Rsun$, and a canonical explosion energy
$E=1.2\times10^{51}$~ergs
\citep{tom09b}. Also, a rapid rise detected by the {\it Kepler}
satellite is also reported for \SNkepler\ ($z=0.087$, \citealt{gar16})
but the detection has been questioned by \cite{rub17}. 
Recently, the rising part of the
shock breakout is firmly detected for the Type IIb \SNmelina\ by a fortunate
amateur astronomer, and exhibits a fast rise rate of
$43\pm6$~mag~day$^{-1}$ \citep{ber18}. A fast rise rate of
$31.2$~mag~day$^{-1}$ is also found in a broad-lined stripped-envelope
supernova SN2018gep, and is interpreted as a shock breakout in a massive
shell of dense circumstellar medium (CSM, \citealt{ho19}).

On the other hand, the UV emission of \SNps\ is
$\sim1$~mag brighter than that of \SNsnls\ and its rise
time and duration are $\gsim50$~times longer than the radiative diffusion
time and the light-crossing time of the shock breakout at the stellar
surface \citep{gez15}, and thus the UV burst is interpreted as a
shock breakout in a circumstellar wind with high mass loss rate
of $10^{-3}~\Msun~{\rm yr^{-1}}$.
The shock breakout in a dense wind is also proposed for the Type IIn SN
PTF~09uj \citep{ofe10}. Its NUV light curve rises with a timescale of a few days to
an absolute magnitude of $\sim-19.5$~mag. Its
peak brightness and rapid rise 
have been attributed to the shock breakout in a
dense circumstellar wind with a high mass loss rate of
$\sim10^{-1}~\Msun~{\rm yr^{-1}}$.

Furthermore, immediate optical
follow-up spectroscopic observations exhibit flash-ionized signatures
\citep{gal14} and reveal that the Type II plateau SN SN~2013fs is
surrounded by a dense CSM (\citealt{yar17}). The signature
of dense CSM might appear in the rapid rise in SNe~IIP \citep{mor16},
and the variation of the rising timescale might be explained by 
variations in
the dense CSM \citep{mor17,moriya18}. On the other hand, the existence of 
dense CSM is inconsistent with the UV emission of \SNsnls. \cite{for18}
performed light curve fitting of the SNe~II LC discovered by the HiTS
with theoretical models \citep{moriya18} and found that 24 of 26 SNe~II have
a rapid rise that can be explained by dense CSM with a mass loss rate
of $>10^{-4}~\Msun~{\rm yr^{-1}}$. These observations imply variations in
the final phases of stellar evolution.  High-cadence observations are
needed to reveal the final stages of these massive stars.

Therefore, we performed a high-cadence transient survey with 
Hyper Suprime-Cam (HSC,
\citealt{miy06,miy12})\footnote{{\url
http://www.subarutelescope.org/Observing/Instruments/HSC/index.html}} 
with a field of view of $1.77~{\rm deg^2}$. This is the most
powerful instrument, with the highest light-collecting power per unit
time, currently available to detect rapid transients. We also conducted
follow-up observation as part of a Subaru HSC survey Optimized for
Optical Transients (SHOOT). In this paper, we focus on a rapidly declining
transient named \SNa\ found in the SHOOT Jul 2014 run; rapidly rising
transients from that run are summarized in \cite{tanaka16}.

  \begin{deluxetable*}{ccccccccccc}
  \tabletypesize{\scriptsize}
  \tablecaption{Subaru observations and light curves of \SNa. \label{tab:lc}}
  \tablewidth{0pt}
  \tablehead{
  \colhead{UT} & \colhead{MJD} & \colhead{Epoch} & \colhead{Instrument} & \colhead{Filter} & \colhead{Exposure time} & \colhead{Flux\tablenotemark{a}} & \colhead{Flux error\tablenotemark{a}} & \colhead{AB magnitude}  & \colhead{Significance} & \colhead{Seeing\tablenotemark{b}} \\  
  \colhead{} & \colhead{} & \colhead{} & \colhead{} & \colhead{} & \colhead{[s]} & \colhead{[$\mu$Jy]} & \colhead{[$\mu$Jy]} & \colhead{[mag]} & \colhead{[$\sigma$]} & \colhead{[arcsec]} } \startdata 
  2014-07-02 & 56840.554 & Day~1 & HSC & $g$ &  600 & $0.289$ & 0.042 & $25.25^{+0.17}_{-0.15}$ & $ 7.0$ & 0.566 \\
  2014-07-02 & 56840.591 & Day~1 & HSC & $g$ &  600 & $0.277$ & 0.041 & $25.30^{+0.18}_{-0.15}$ & $ 6.7$ & 0.485 \\
  2014-07-02 & 56840.610 & Day~1 & HSC & $g$ &  600 & $0.358$ & 0.043 & $25.02^{+0.14}_{-0.12}$ & $ 8.4$ & 0.523 \\
  2014-07-03 & 56841.525 & Day~2 & HSC & $g$ &  600 & $0.056$ & 0.043 & $>26.11$\tablenotemark{d} & $ 1.3$ & 0.665 \\
  2014-07-03 & 56841.559 & Day~2 & HSC & $g$ &  600 & $0.087$ & 0.044 & $>26.09$\tablenotemark{d} & $ 2.0$ & 0.581 \\
  2014-07-03 & 56841.596 & Day~2 & HSC & $g$ &  600 & $0.118$ & 0.040 & $26.22^{+0.45}_{-0.32}$ & $ 3.0$ & 0.552 \\
  2014-07-03 & 56841.615 & Day~2 & HSC & $g$ &  600 & $0.095$ & 0.057 & $>25.82$\tablenotemark{d} & $ 1.7$ & 0.600 \\
  2014-08-05 & 56874.339 & Day~35 & FOCAS & $g$ &  960 & $0.117$ & 0.090 & $>25.32$\tablenotemark{d} & $ 1.3$ & 0.872 \\
  2015-05-24 & 57166.513 & Day~327& HSC & $g$ & 960 & ---\tablenotemark{c}  & ---\tablenotemark{c} & ---\tablenotemark{c} & ---\tablenotemark{c} & 0.950 \\ \hline
  2014-07-02 & 56840.585 & Day~1 & HSC & $g$ & 1800 & $0.296$ & 0.036 & $25.22^{+0.14}_{-0.13}$ & $ 8.1$ & 0.523 \\
  2014-07-03 & 56841.574 & Day~2 & HSC & $g$ & 2400 & $0.092$ & 0.037 & $26.50^{+0.56}_{-0.37}$ & $ 2.5$ & 0.593 \\ \hline
  2014-07-02 & 56840.467 & Day~1 & HSC & $r$ &  600 & $0.214$ & 0.052 & $25.57^{+0.30}_{-0.23}$ & $ 4.2$ & 0.799 \\
  2014-07-03 & 56841.445 & Day~2 & HSC & $r$ &  600 & $0.092$ & 0.065 & $>25.68$\tablenotemark{d} & $ 1.4$ & 0.545 \\
  2014-08-05 & 56874.326 & Day~35 & FOCAS & $r$ &  960 & $0.052$ & 0.098 & $>25.23$\tablenotemark{d} & $ 0.5$ & 0.950 \\
  2015-08-19 & 57253.501 & Day~414& HSC & $r$ & 1440 & ---\tablenotemark{c}  & ---\tablenotemark{c} & ---\tablenotemark{c} & ---\tablenotemark{c} & 1.410 
  \enddata
  \tablenotetext{a}{Corrected for the Galactic extinction with the color
  excess of $\Ebvg=0.036$~mag. }
  \tablenotetext{b}{Full width at half maximum.}
  \tablenotetext{c}{Used for the reference image.}
  \tablenotetext{d}{Calculated from the $3\sigma$ error.}
  \end{deluxetable*}
 
This paper consists of following sections. In Section~\ref{sec:obs}, the
observations are described. In Section~\ref{sec:result}, observational
properties of \SNa\ are summarized. In Sections~\ref{sec:other} and
\ref{sec:theoretical}, these properties of \SNa\ are compared with those
of known transients and variables, and of theoretical models,
respectively. In Section \ref{sec:conclusion}, we discuss the
findings and
present our conclusions. In this paper, we adopt the AB magnitude
system and the {\it WMAP5} cosmological
parameters: $H_0=70.5~{\rm km~s^{-1}~Mpc^{-1}}$, $k=0$,
$\Omega_\lambda=0.726$, and $\Omega_{\rm M}=0.273$ \citep{kom09}.

  \begin{deluxetable*}{ccccccccccc}
  \tabletypesize{\scriptsize}
  \tablecaption{Imaging observations and photometry of the host galaxy
   of \SNa. \label{tab:host}}
  \tablewidth{0pt}
  \tablehead{
  \colhead{UT} & \colhead{Epoch} & \colhead{Instrument} &
   \colhead{Filter} & \colhead{Exposure time} & \colhead{CModel
   Flux\tablenotemark{a}} & \colhead{Flux error\tablenotemark{a}} &
   \colhead{Seeing\tablenotemark{b}} \\
  \colhead{} & \colhead{} & \colhead{} & \colhead{} &
   \colhead{[s]} & \colhead{[$\mu$Jy]} & \colhead{[$\mu$Jy]} &
   \colhead{[arcsec]} } \startdata
  2015-05-24 & Day~327& HSC & $g$ & 960 & 0.473  & 0.028 & 0.950 \\
  2015-08-19 & Day~414& HSC & $r$ & 1440 & 0.743  & 0.038 & 1.410 \\ \hline
  2016-05-07 & Day~676& S-Cam & $r'$ & 1500 & 0.678  & 0.036 & 1.102 \\
  2016-05-07 & Day~676& S-Cam & $i'$ & 1500 & 0.814  & 0.047 & 0.778 \\
  2016-05-07 & Day~676& S-Cam & $Y$ & 3600 & 0.902  & 0.230 & 0.809
  \enddata
  \tablenotetext{a}{Uncorrected for the Galactic extinction with the color
  excess of $\Ebvg=0.036$~mag. }
  \tablenotetext{b}{Full width at half maximum.}
  \end{deluxetable*}

\section{Observations and data reduction}
\label{sec:obs}

\subsection{Imaging observations of transients}
\label{sec:image}

\begin{figure*}
\plotone{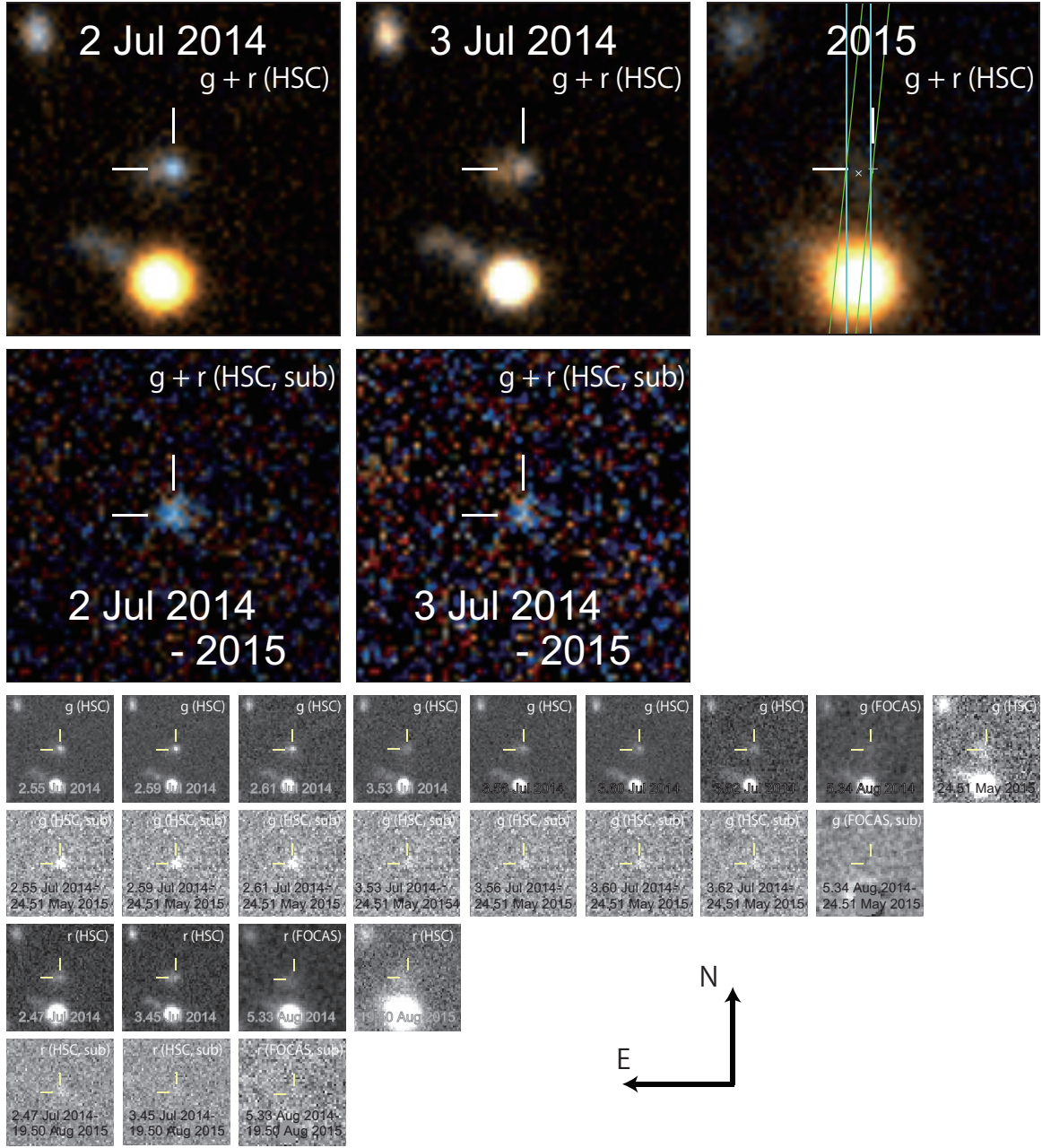} 
 \caption{Images at the location of \SNa\ (ticks). \SNa\
 is located at $0.\hspace{-.5ex}''44$ west and
 $0.\hspace{-.5ex}''14$ north of the host galaxy with the full width at
 half maximum size of
 $1.\hspace{-.5ex}''10$ in the $r$ band. In the multicolor image
 taken in 2015, the positions of \SNa\ and the
 center of the host galaxy are shown with plus and cross, respectively,
 and the slits for spectroscopic observations with FOCAS and GMOS are
 shown with green and cyan boxes, respectively. The slit positions are
 centered at the host galaxy and the position angle are $-13.3^\circ$
 east of north for the FOCAS observation and $0^\circ$ east of north for
 the GMOS observation. The lengths of ticks are $1$~arcsec and the
 figure size is $10\times10$~arcsec$^2$.
}
\label{fig:cutout}
\end{figure*}

The $g$- and $r$-band imaging observations were carried out on 7 fields 
with HSC on 2 and 3 Jul 2014 (UT) (Days 1 and 2). In this paper, we
focus on a candidate \SNa\ at \RADECa\ found in one of our survey
fields (Figure~\ref{fig:cutout}). The field was observed three times in the $g$ band and once in
the $r$ band on Day 1, and four times in the $g$ band and once in the $r$
band on Day 2. Each exposure unit
on Days 1 and 2 consists of five $2$~min frames with
ditherings, and the time intervals between the exposure units are about an hour. 
We also took $g$- and $r$-band images of $16$~min exposures with Subaru Faint Object
Camera And Spectrograph (FOCAS; \citealt{kas02}) on 5 Aug 2014 (Day 35), a $g$-band image of $16$~min exposure
with HSC on 24 May 2015 (UT) (Day 327), and an $r$-band image of
$24$~min exposure with HSC on 19 Aug 2015 (UT) (Day 414). The details of
the imaging observations of transients are summarized in Table~\ref{tab:lc}.

For our transient search, the HSC data are reduced using HSC pipeline\footnote{A prototype is
described in \cite{fur10}.} \citep{bos18} version 3.6.1, which is 
based on the LSST pipeline \citep{ive08,axe10}. It provides packages
for bias subtraction, flat fielding, astrometry, flux calibration, mosaicing, warping,
coadding, and image subtraction. The astrometry and photometry are made
relative to the Sloan Digital Sky Survey (SDSS) DR8 \citep{aih11} with a
$2.36$~arcsec ($14$~pixel) diameter aperture. We 
developed a quick image subtraction system with the HSC pipeline and
performed realtime transient finding \citep{tom14atel}, in cooperation with
an on-site data analysis system (\citealt{fur11,fur18}). This enables us
to report
numerous SN candidates immediately after the observing runs
\citep{tom14atel,tom14atel2,tom15atel,tom15atel2}, and to search for optical
counterparts of gravitational waves and fast radio bursts
\citep{uts18gw151226,tom18frb,tom18gw170817}. On the other hand, the
image subtractions between the HSC images and the FOCAS images are made
with {\sc HOTPANTS} \citep{bec15} after 
warping the HSC images with {\sc WCSRemap}. For these image
subtractions, the difference imaging method of Alard \& Lupton
\citep{ala98,ala00} is adopted, which handles point-spread
function (PSF) variations and allows successful image
subtraction between images with different PSF sizes. 

In order to detect candidates with short time variabilities, the first
$g$- and $r$-band images obtained on Day 1 are set as reference images
for the image subtraction. Objects with more than one
detection at significance higher than $5\sigma$ in the difference
images are selected as rapid transient candidates. In other words, the
candidates show variability between observations on 
two successive nights. On the other hand, for the
photometry, we adopt images taken on Day 327 and Day 414 as reference
images for the $g$ and $r$ bands, respectively, and subtract the
reference images from hourly-stacked and daily-stacked images 
of Days 1 and 2.
We evaluate astrometric accuracy in the difference images by a random
injection of artificial point sources to images before
warping, coadding, and image subtraction. After warping, coadding, image
subraction, and source detection, the positions of artificial sources
with $25-25.5$~mag are recovered to a median accuracy of $<0.14$~arcsec.
The fluxes are measured with aperture photometry using a
$2.36$~arcsec-diameter aperture on the difference images (Figure~\ref{fig:lc}). 

\subsection{Imaging observations of the host galaxy}
\label{sec:imageHost}

\begin{figure}
\plotone{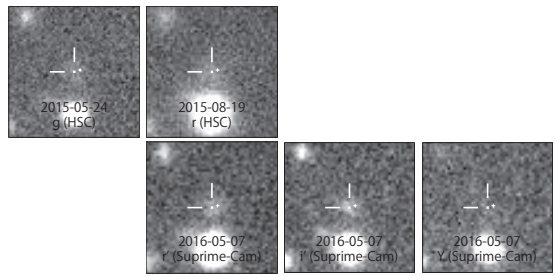} 
 \caption{Multicolor images of the host galaxy. The lengths of ticks
 are $1$~arcsec and the figure size is $10\times10$~arcsec$^2$. The positions of \SNa\ and the
 center of the host galaxy are shown with plus and cross, respectively.
}
\label{fig:host}
\end{figure}

In addition to the $g$-band observation with HSC on Day 327 and
the $r$-band observation with HSC on Day 414, we performed
$r'$-, $i'$-, and $Y$-band imaging observations with Suprime-Cam
\citep{miy02} on 7 May 2016 (UT) (Day 676), in order to derive
properties of a host galaxy at \RADEChost\ (Figure~\ref{fig:host}). 
For the multi-band photometry of the host
galaxy, we adopt HSC pipeline version 4.0.1 to derive consistently
composite model (CModel) fluxes with exponential and de Vaucouleurs
fits.\footnote{{\url http://www.sdss.org/dr12/algorithms/magnitudes/\#cmodel}} The imaging observations
and photometry of a host galaxy are summarized in Table~\ref{tab:host}.
As the pixel scales of Suprime-Cam and HSC are
different, the multi-band data analysis is separately performed for
Suprime-Cam and HSC images. The discrepancy between the $r$-band flux
with an HSC image and the $r'$-band flux with a Suprime-Cam image is
taken into account as a systematic error in the subsequent analysis.
The position of host galaxy is also derived from the multi-band
analyses \citep{bos18}. The position in each image is measured by an approximate
maximum-likelihood algorithm implemented in the HSC pipeline
(\citealt{bos18}, see also \citealt{pie03}). The method gives a median
root-mean-square of $<0.1$~arcsec at a seeing of $<2$~arcsec in the SDSS
with a pixel scale of $0.396$~arcsec \citep{pie03}.
We adopt the average position of the host galaxy derived by the
multi-band analyses of the HSC images and the Suprime-Cam images.
These multi-band analyses give a consistent position of
the host galaxy within $0.14$~arcsec.

\subsection{Spectroscopic observation}
\label{sec:spectra}

Optical spectroscopic observations were carried out 
with FOCAS on the 8.2-m Subaru telescope on 5 August 2014 (Day 35) and 
with the Gemini Multi-Object Spectrograph (GMOS) \citep{hoo04} on the 8.1-m Gemini-North telescope on 10 and 11 Jun 2016
(Days 710 and 711).

In the FOCAS observation, we aimed to take a spectrum of the transient and took 
four $20$-minute exposures. Low-resolution spectra
($R\sim600$) were obtained in the multi-object slit mode with
the $0.\hspace{-.5ex}''8$-width slit, 300 lines mm$^{-1}$ blue (300B)
grism, and the SY47 order-sort filter, providing wavelength coverage
of $4700$~\AA-$9000$~\AA. We adopted the $2\times1$ binning
mode in spatial and spectral directions, respectively. 
The obtained spatial and spectral samplings are
0.208~arcsec pixel$^{-1}$ and $1.4$~\AA~pixel$^{-1}$, respectively.
A standard star, BD+28d4211, was observed on 
the same night in the same observing mode and is used for flux calibration.
The data are reduced with the {\sc IRAF} packages \citep{tod86,tod93} in a standard manner.

In addition, we took four $22.5$-minute exposures with GMOS 
to take the host galaxy spectrum 
on 10 and 11 June, 2016. 
A medium-resolution spectrum ($R\sim3000$) was
obtained in the long slit mode with the $0.\hspace{-.5ex}''75$-width
slit, 831 lines mm$^{-1}$ (R831\_G5302) grism, and the GG455\_G0305
order-sort filter, providing the wavelength coverage of
$5200$~\AA-$7300$~\AA. We adopted the $2\times2$ binning mode and the
obtained spatial and spectral samplings are 0.145~arcsec pixel$^{-1}$
and $0.67$~\AA~pixel$^{-1}$, respectively. We stacked the four spectra
after removing cosmic rays from each 2-dimensional spectrum.
We use a standard star, Feige~34, for flux calibration,
based on observations taken
on 24 April 2016 
in a different configuration, with the same grism and no order-sort filter. 
The effects on our flux calibration of 
these different configurations for the target and the standard star 
are evaluated in the next paragraph.
The data are reduced with
the {\sc Gemini-IRAF} reduction software package in a standard manner.

The slit of the GMOS spectrum as well as that of the FOCAS spectrum 
was aligned with a nearby bright star, SDSS~J213304.29+093551.6. 
The SDSS $r$-band magnitude is $r=20.60\pm0.03$ according to SDSS DR14. 
The GMOS spectrum of this star is well matched with an M1-type star. 
An apparent $r$-band magnitude of the star obtained by convolving the observed spectrum 
with the SDSS $r$-band response function is 21.16~mag and the difference of the magnitudes is 0.56~mag, 
corresponding to a factor of 1.67 in brightness.
This is attributed to the different order-sorting filters between the
target and standard star, slit losses, and atmospheric extinction.
As observed line fluxes of the host galaxy measured in \S\ref{sec:host}
are likely to be lost as the flux of SDSS~J213304.29+093551.6 does,
they are multiplied by a factor of 1.67.

\section{Observational properties}
\label{sec:result}

\begin{figure}
\plotone{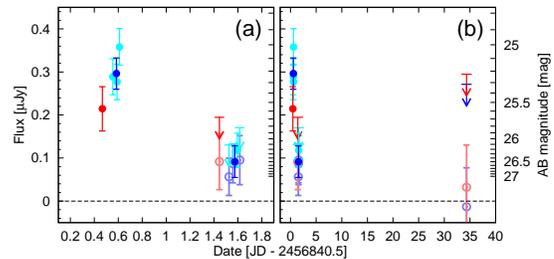} 
\caption{Multicolor light
 curves of \SNa\ (cyan, $g$ band [hourly-stacked]; blue, $g$ band
 [daily-stacked]; red, $r$ band) and the redshifted near
 UV light curve of
 \SNsnls\ (gray) (a) at the shock breakout phase and (b) from the shock
 breakout to $\sim$~month after the explosion. The $3\sigma$
 upper limits and the open symbols are shown at the phase
 when the significance is low ($\leq2\sigma$).
}
\label{fig:lc}
\end{figure}

\subsection{\SNa}
\label{sec:SHOOT14di}

Figure~\ref{fig:cutout} shows the multicolor images constructed from
the $g$- and $r$-band images taken with HSC, the monochromatic $g$- and
$r$-band images taken with HSC and FOCAS, and the difference images
after the image subtraction.
An object with decline between Days 1 and 2 is found in the $g$ band at
$0.\hspace{-.5ex}''44$ west and $0.\hspace{-.5ex}''14$ north of the host
galaxy. The full width at half maximum size of the host galaxy is
$1.\hspace{-.5ex}''10$ in the $r$ band image taken on Day~2.

The $g$-band fluxes on Days 1 and 2 are $\sim0.28-0.36$ and
$\sim0.1~\Junit$, respectively. The flux measurement with the $g$-band
daily-stacked images also gives consistent results. The $g$-band flux
declines at a rate of $-0.204\pm0.052~\Runit$, corresponding to
$+1.28^{+0.40}_{-0.27}~\Munit$ in the observer frame between Days~1 and 2. Here, the
error is evaluated with a geometric mean of the $1\sigma$ flux errors.
The $r$-band flux on Day 1 is $0.21~\Junit$. On the other hand,
nothing is significantly detected in the difference image between the
$r$-band images taken on Days 2 and 414, or in the difference image
between the FOCAS and HSC images. The $3\sigma$ upper limits on Day 35 are
$0.27~\Junit$ in the $g$ band and $0.30~\Junit$ in the $r$ band
(using a $2.36$~arcsec diameter aperture). The fluxes are summarized
in Table~\ref{tab:lc}, and the $g$- and $r$-band light curves are shown
in Figures~\ref{fig:lc}(a) and \ref{fig:lc}(b). Here, the
Galactic reddening is corrected with a color excess $\Ebvg$ of $0.036$~mag
\citep{sch11}\footnote{\url{http://irsa.ipac.caltech.edu/applications/DUST/}}
and a fitting function of the Galactic extinction curve \citep{pei92}. 

In order to characterize time variability, we introduce a rate of
brightness change $R(\lres,\Delta t_{\rm rest})$ at a wavelength $\lres$
within an interval $\Delta t_{\rm rest}$ in the rest frame. For \SNa\ at
$z=0.42285$ (Section~\ref{sec:host}),
$1$~day in the observer frame corresponds to
$\Delta t_{\rm rest}=0.70$~day, and the effective rest wavelength is
$3400$~\AA\ for the $g$ band. The
observations illustrate that $\Rgd$ of \SNa\ between Days 1 and 2 is
$-0.294\pm0.074~\Runit$. 
The rate at $\lres=3400$~\AA\ is also written in terms
of magnitude as $\Rgd=+1.83^{+0.57}_{-0.39}~\Munit$.

\begin{figure}
 \plotone{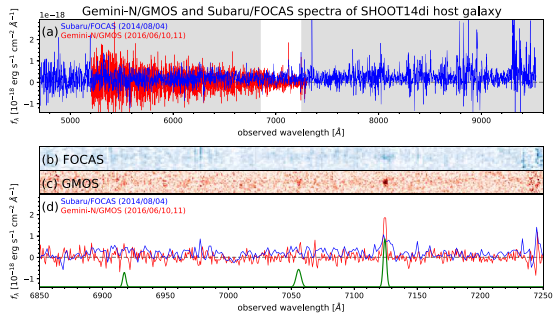}
 \caption{
 Observed spectra of the host galaxy of SHOOT14di with 
Gemini-North GMOS (red) and Subaru FOCAS (blue). 
Best-fitted gaussian for the H$\beta$ and two [OIII] lines are shown in green 
although the detection of H$\beta$ line is marginal. 
(a): the spectra in almost the entire range of the FOCAS spectrum, 
(b): a magnified view of the 1d spectra around the detected emission lines, 
(c), (d): magnified views of the two-dimensional GMOS and FOCAS spectra. 
}
\label{fig:spectra}
\end{figure}

\begin{figure*}
\plotone{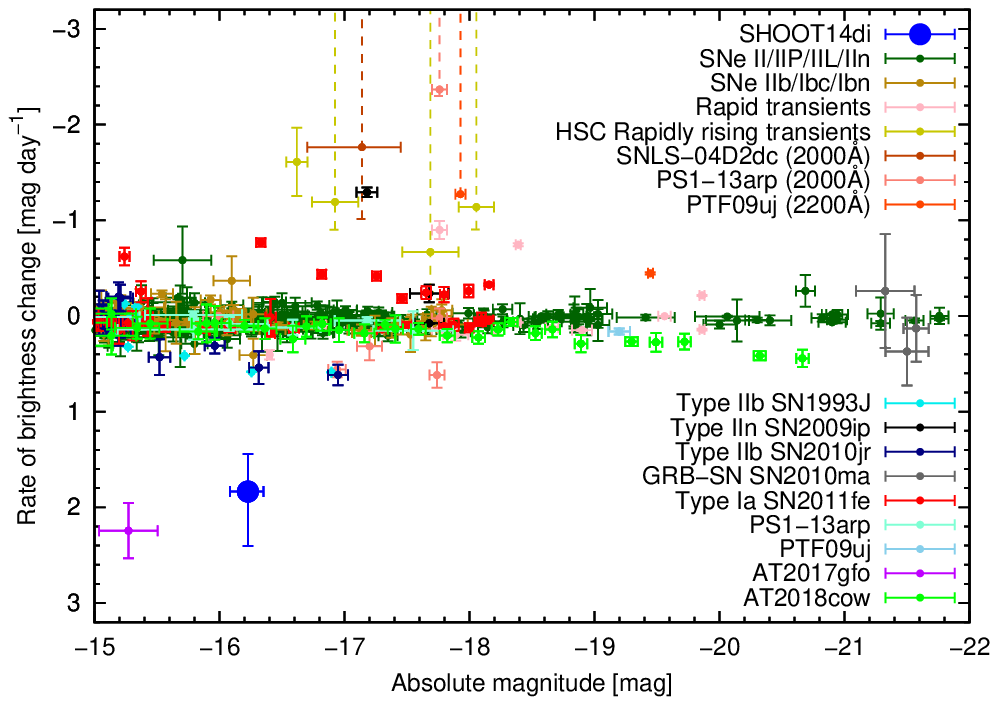}
\caption{Absolute magnitudes and rates of brightness change of \SNa\
 (blue) compared with those of shock breakouts (\SNsnls\
 [dark-orange] \citealt{sch08,gez08}, PS1-13arp [salmon]
 \citealt{gez15}, and PTF~09uj [orange-red]
 \citealt{ofe10}), other SNe (Type Ia SN~2011fe [red]
 \citealt{nug11,bro12}; Type IIb SN~1993J [cyan] \citealt{ric94,ric96}; Type
 IIn SN~2009ip [black], Type IIb SN~2010jr [navy], Type II/IIP/IIL/IIn
 SNe [dark-green], Type IIb/Ibc/Ibn SNe [dark-goldenrod] \citealt{pri14}),
 rapid transients \citep[light-pink,][]{dro14}, rapidly rising
 transients \citep[dark-yellow,][]{tanaka16}, AT2017gfo
 \citep[dark-magenta,][and references therein]{GW170817OptSum}, and AT2018cow
 \citep[green,][]{pre18,per19}. The absolute magnitudes
 and rates of brightness change of PS1-13arp [aquamarine] and PTF~09uj
 [skyblue] at the phases other than the shock breakout are also shown. These values are
 $K$-corrected to $\lambda_{\rm rest}=3400$~\AA, except for the shock
 breakout of \SNsnls, PS1-13arp, and PTF~09uj. The data of PS1-13arp on $MJD=56401.01$ is
 excluded from the analysis because of the short exposure time
 (S.~Gezari, private communication).
}
\label{fig:dmag}
\end{figure*}

\subsection{Host galaxy}
\label{sec:host}

The FOCAS spectrum taken on Day 35 does not show any SN features, 
which is reasonable 
considering the depth of the spectroscopic observations and the faintness of the transient on that day. 
We detected an emission line on the faint continuum
from the host galaxy at $7124.4$~\AA. 
The second spectrum taken with GMOS focused on the wavelength region around this single emission line. 
Thanks to the higher resolution of the GMOS spectrum, 
we successfully detected another emission line
at $7056.2$~\AA\ as shown in Figure~\ref{fig:spectra}.  
These two emission lines correspond to the
\ion{[O}{3]} doublet\ $\lambda\lambda5007,4959$ at $z=0.4229$.
The $gr$ HSC and $r'i'Y$ Suprime-Cam broad-band photometry is fitted by
a Bayesian photometric-redshift code \citep[][]{tanaka15photoz} with the redshift fixed
to the spectroscopic redshift to infer the stellar mass of the galaxy.
Using the \citet{bur03} stellar population synthesis models
for the Chabrier initial mass function \citep{cha03}, we find
that the galaxy is a low-mass galaxy with $M_*=1.2\times10^8\rm M_\odot$.
The code applies priors on the physical properties of galaxies such as star
formation rate, stellar mass, age, and dust extinction, but the inferred
stellar mass does not significantly change if we disable all the priors.
The typical statistical uncertainty in the stellar mass is 0.2-0.3~dex
when the redshift is fixed \citep[][]{tanaka15photoz}.  This level of uncertainty does not
affect the following discussions.

The combined line flux of the two \ion{[O}{3]} emission
lines is $6.0\times10^{-18}$~erg~s$^{-1}$~cm$^{-2}$. Our detection of the
H$\beta$ emission line is marginal. The gaussian-fitted and $3\sigma$
upper limits of H$\beta$ line flux are $1.4\times10^{-18}$~erg~s$^{-1}$~cm$^{-2}$
and $2.8\times10^{-18}$~erg~s$^{-1}$~cm$^{-2}$, respectively.
The \ion{[O}{2]} emission line at this redshift is not detected and 
only a weak upper limit of 
$1.0\times10^{-17}$~erg~s$^{-1}$~cm$^{-2}$ ($3\sigma$) can be set.
The lower limit of the flux ratio of
$\log_{10}\rm{\{f([OIII])/f(H\beta)\}}=+0.82$ roughly corresponds to
a metallicity of $12+\log_{10}\rm{(O/H)}\sim7.7-8.3$ \citep{nagao06}.
Although [OIII] emission lines are not a good proxy for measuring the star
formation rate of a galaxy in general, we use them here to put a constraint on
the star formation rate of the host galaxy.
We adopt a typical $\log_{10}{(\rm{[OIII]/H}\beta)}$ line flux ratio of
$\sim+0.6$ \citep{jun14} to convert the [OIII] flux to H$\beta$ flux,
which is roughly consistent with our measurements.
Given the small stellar mass derived from the
photometric-redshift code, we expect the dust extinction to be
small or almost zero
\citep{gar10}. Then, assuming case B
recombination and zero extinction, we obtain an H$\alpha$ line luminosity of
$3.8\times10^{39}$~erg~s$^{-1}$ and star formation rate of 0.03~$\Msun$~yr$^{-1}$
\citep{ken94}.
Taking into account the extended nature of the host galaxy inferred
from the difference between its Kron and PSF magnitude, its absolute
flux density could be about three times larger than the current
estimate, resulting in the star formation rate of $\sim0.1$~$\Msun$~yr$^{-1}$.
A ratio of the faint H$\alpha$ emission line luminosity and
faint absolute broad-band magnitude roughly follows previous results
on H$\alpha$ emitting galaxies \citep{tre02}.

 \section{Comparisons with known transients and variables}
\label{sec:other}

\subsection{Supernovae, rapid transients, and a kilonova}
\label{sec:otherSN}

\begin{figure}
\plotone{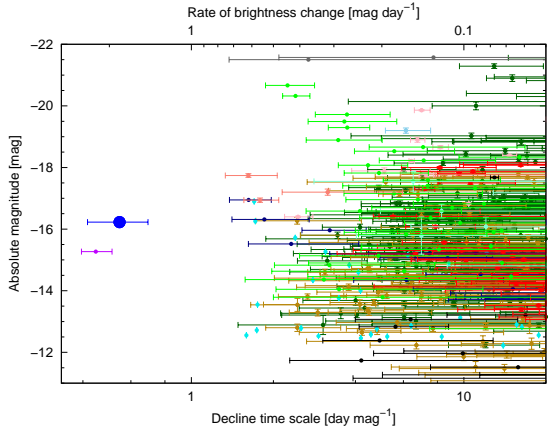}
\caption{Decline time scales and absolute magnitudes of \SNa, other SNe, rapid
 transients, rapidly rising transients, and a kilonova. The colors of points are the same as
 Figure~\ref{fig:dmag}.
}
\label{fig:timescale}
\end{figure}

\begin{figure}
\plotone{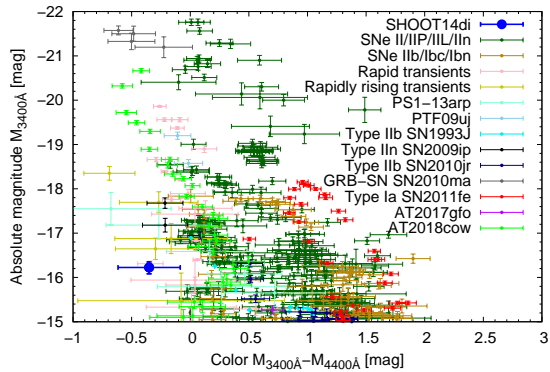}
\caption{Color magnitude diagram of \SNa\ (blue), other SNe, rapid
 transients, rapidly rising transients, and a kilonova, except for the shock
 breakout of \SNsnls, PS1-13arp, and PTF~09uj. The colors of points are the same as
 Figure~\ref{fig:dmag}.}
\label{fig:color}
\end{figure}

\begin{figure}
\plotone{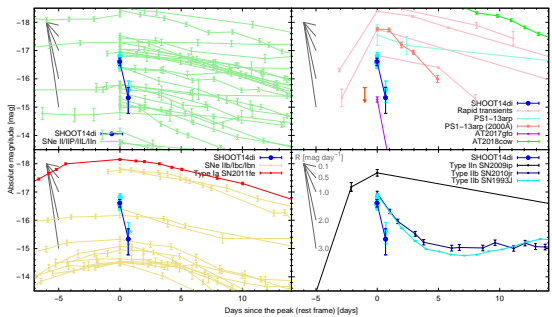}
\caption{Light curve of \SNa\ (blue) compared with those of SNe~II/IIP/IIL/IIn
 (top left), rapid transients, PS1-13arp, AT2017gfo, and AT2018cow (top right), SNe~Ia/IIb/Ibc/Ibn (bottom
 left), and Type IIn SN~2009ip, Type IIb SN~2010jr, and Type IIb
 SN~1993J (bottom right).
}
\label{fig:lccomp}
\end{figure}

The peak magnitude, time variability, and color of \SNa\ are compared
with SNe, rapid transients recently
 pointed out by \cite{dro14}, and a kilonova found in 2017 \citep{GW170817MMApaper}. Figure~\ref{fig:dmag} shows
the rate of brightness change at $\lambda_{\rm rest}\sim3400$~\AA\ as a function of the absolute magnitude 
(based on the brighter of the two observations used for
the rate estimate), while Figure~\ref{fig:timescale} shows
the absolute magnitude at $\lambda_{\rm rest}\sim3400$~\AA\ as a function of
the time scale of the variability. Figure~\ref{fig:color} is a color-magnitude diagram
with an absolute magnitude $\Mg$ at $\lambda_{\rm rest}\sim3400$~\AA\
and a color, $\Mg-\Mr$, derived with magnitudes at
$\lambda_{\rm rest}\sim3400$ and $\sim4400$~\AA. Figure~\ref{fig:lccomp}
shows comparisons of light curves at $\lambda_{\rm rest}\sim3400$~\AA\
with the other transients. The rate and color-magnitude diagram of other
SNe and transients is derived
from the $u$- and $b$-band light curves of nearby SNe obtained by the
{\it Swift} Ultraviolet/Optical Telescope (UVOT)
\citep{nug11,bro12,pri14}, the $U$- and $B$-band light curves of
SN~1993J \citep{ric94,ric96}, the $g$- and $r$-band light curves of
PTF~09uj after the shock breakout obtained by {\it PTF} \citep{ofe10},
the $g$- and $r$-band light curves of \SNps\ after the
shock breakout and rapid transients obtained by {\it Pan-STARRS1}
\citep{gez15,dro14}, the $g$- and $r$-band light curves of the rapidly
rising transients obtained by {\it HSC} \citep{tanaka16}, $uvw1$-,
$u$-, and $g$-band light curves of 
a fast luminous ultraviolet transient AT2018cow \citep{pre18,per19}, and
$U$-, $g$-,
and $r$-band light curves of a kilonova (AT2017gfo) associated with a
gravitational wave source GW170817
(\citealt{GW170817andreoni,GW170817arcavi,GW170817coulter,GW170817cowperthwaite,GW170817diaz,GW170817drout,GW170817evans,GW170817valenti,GW170817mansi,GW170817pian,GW170817shappee,GW170817smartt,utsumi17},
summarized in \citealt{GW170817OptSum}). We
apply $K$-corrections by interpolating or
extrapolating their spectral energy distributions in magnitude. We also
calculate the lower limits of the rise rates of \SNsnls, PTF~09uj, and \SNps\
with $3\sigma$ upper limits at $1$~day or $2$~days before their detection and the
rate of brightness change of PTF~09uj and \SNps\ from their NUV light
curves obtained with the {\it GALEX} satellite;
however, $K$-corrections
are not applied to the NUV light curves of \SNsnls, PTF~09uj, and
\SNps\ because there are no simultaneous observations with high
signal-to-noise ratios in other bands.

Figures~\ref{fig:dmag}, \ref{fig:timescale}, \ref{fig:color}, and
\ref{fig:lccomp} illustrate following characteristics of \SNa.
\begin{itemize}
 \item All of SNe and SLSNe decline more slowly than \SNa. The colors of
       most of them are redder than that of \SNa\
       ($\Mg-\Mr\sim-0.4$). The transient most similar to \SNa\ is the cooling
       tail of the shock breakout of Type IIb supernovae. However, its
       brightness change is half as fast as that of \SNa.
 \item The colors of the GRB-associated Type Ic SN~2010ma and the shock
       breakout in the dense circumstellar wind of PTF~09uj are similar to
       that of \SNa. The absolute rate of brightness change of PTF~09uj
       at the first NUV detection is consistent
       with that of \SNa. However, their brightness is brighter than
       that of \SNa\ and the rate of brightness change of PTF~09uj is
       slower than that of \SNa\ at the same brightness.
 \item The colors of rapid transients at their peak are as
       blue as that of \SNa\ and their brightness is also consistent
       with \SNa. However, their decline is slow in spite of their
       rapid rise. \cite{dro14} indicate that their decline
       timescales $t_{1/2,{\rm decline}}$ are
       $t_{1/2,{\rm decline}}\gsim2$~days, corresponding to
       $\Rd\lsim+0.26~\Munit$, which is slower than
       \SNa.
 \item The lower limit of the temperature of \SNps\ at its earliest
       epoch ($T>2.2\times10^4$~K) indicates a blue color of \SNps\
       ($\Mg-\Mr<-0.11$~mag), which is consistent with that of \SNa. The
       color and brightness of \SNps\ just after the shock breakout are
       also consistent with that of \SNa, and the absolute rate of
       brightness change of \SNps\ at the first NUV detection is consistent
       with that of \SNa. However, the decline rate of \SNps\
       just after the peak, even in the NUV band,
       ($\Rdps=+0.62\pm0.13~\Munit$) is slower than that of \SNa\ in the
       optical bands.
 \item The absolute rate of brightness change of \SNsnls\ at the first
       NUV detection is consistent with that of \SNa. Unfortunately, the
       decline rate of \SNsnls\ cannot be well constrained owing to its
       low signal-to-noise ratios.
 \item The absolute rate of brightness change and color of
       rapidly rising transients reported in \cite{tanaka16} are similar
       to those of \SNa. The origin of the rapid rise is suggested to be
       emission by a cooling envelope or the shock breakout in a dense
       circumstellar wind, depending on the brightness and the rise
       rate.  Unfortunately, there are no observation of their decline.
 \item While the color of AT2018cow at its peak is consistent with that of
       \SNa, AT2018cow is redder than \SNa\ at the same brightness. The
       rates of brightness change of AT2018cow are $\Rgd\sim+0.44~\Munit$ at the
       peak and $\Rgd\sim+0.15~\Munit$ at $\Mg=-16$. Both rates are
       slower than that of \SNa.
 \item The brightness change of \SNa\ is as rapid as that of
       AT2017gfo. However, the peak magnitude and color of \SNa\ is $1$~mag
       brighter and $1$~mag bluer than those of AT2017gfo,
       respectively. 
\end{itemize}

The discovery of the rapidly-declining\\
Type I SN~2019bkc/ATLAS19dqr has
recently been reported \citep{che19}. In the observer frame, SN~2019bkc
exhibits slower decline rates, $+0.45-+0.63~\Munit$ in the $B$-,
$V$-, $g$-, $r$-, and $i$-bands, and a redder color at the peak, than
\SNa. Although the absolute brightness, the rates, and the color in the
rest frame of SN~2019bkc cannot be constrained because it appears to be
hostless, the origin of SN~2019bkc is likely to be different from that
of \SNa.

\subsection{Other transients or variables}
\label{sec:otherOther}

There are other transients or variables with time scales shorter than
SNe and similar transients. In this section, we review their
properties and compare them to those of \SNa.

Considering the astrometric accuracy for the positions of artificial
sources in the difference images and the position of host galaxy
(sections~\ref{sec:image} and \ref{sec:imageHost}), \SNa\ appeared at
the outskirt of the host galaxy and thus is unlikely to be an active
galactic nucleus or a tidal disruption event. 
We rule out an asteroid 
because the displacement between Days 1 and 2 is less than $0.17$~arcsec
and no variable, \ie moving, object is found within $1$~arcmin around
\SNa\ on Day 327.

The decline rate of \SNa\ could be reproduced with the optical flash and
afterglow of an on-axis GRB, or the orphan afterglow of an off-axis
GRB. However, \SNa\ is continuously bright over $\geq56$~min in the
rest frame on Day 1, which is longer than the typical duration of 
on-axis GRBs, and there were no alerts of GRBs at the location of \SNa\ on
Days 1 and 2. On the other hand, the duration on Day 1 and the decline
rate between Days 1 and 2 could be explained by a GRB orphan afterglow
peaking at $\sim0.6-1.4$~days, depending on the power-law index $\alpha$
of light curve decline, where $F\propto t^{-\alpha}$. Here we adopt
$\alpha=1.6-3.0$ \citep[\eg][]{tot02,gra02}. The early peak of the
GRB orphan afterglow and the
non-detection of on-axis GRBs require a viewing angle
$\theta_{\rm obs}$ 
of $\theta_{\rm jet} < \theta_{\rm obs} \lsim 7^\circ$
for the opening angle $\theta_{\rm jet}$ of $\theta_{\rm jet}\lsim 4^\circ$,
or $\theta_{\rm jet} < \theta_{\rm obs} \lsim \theta_{\rm jet} + 2^\circ$
for $6^\circ \lsim \theta_{\rm jet}\lsim 14^\circ$.
The solid angle required for GRB orphan afterglows by the early
peak is similar to that for the on-axis GRBs, and thus their occurrence
rates are likely to be comparable. Adopting the cosmic GRB rate \citep{lie14},
the expected occurrence number of GRB orphan afterglows with an early
peak during the HSC observation on Days 1 and 2 is $\lsim 0.001$ for
GRBs at $z\leq2$. The number is much smaller than unity, and thus 
a serendipitous detection is unlikely. Furthermore, the optical
spectrum of the GRB afterglow is proportional to $\nu^{-1/2}$ at
$\sim1$~day \citep{sar98,gra02}, and thus the color of the GRB orphan
afterglow ($\Mg-\Mr\sim+0.14$) is redder than that of \SNa.

\begin{figure*}
\plotone{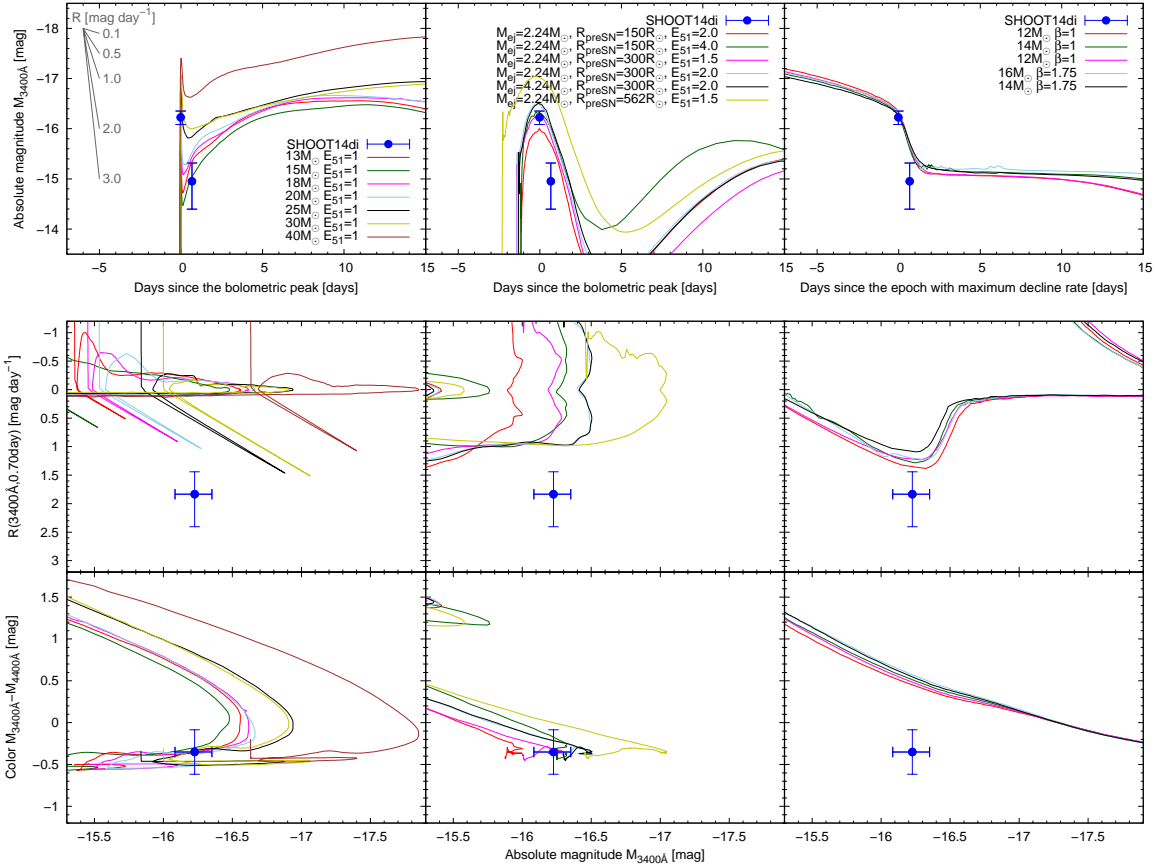}
\caption{Light curve (top), rate of brightness change (middle), and
 color (bottom) of \SNa\ (blue)
 compared with those of theoretical models of
 a shock breakout at the stellar surface of SN~IIP (left), a cooling envelope
 emission of SN~IIb (middle), and an explosion of red supergiant star with a dense
 circumstellar wind with $E_{51}=0.5$,
 the mass loss rate of $10^{-2}~\Msun~{\rm yr}^{-1}$, and the outer radius of dense
 wind of $10^{15}$~cm (right). 
}
\label{fig:model}
\end{figure*}

Another possible candidate is a chance projection of the
flare of a low-mass Galactic star.
Since \SNa\ was detected on Days 1 and 2 and the typical duration of
flare is $\sim20$~min, two flares with durations of $\geq1.34$~hr and
$\geq2.16$~hr are required to explain the decline rate of \SNa. 
The occurrence rate of flares is $\sim2~{\rm hr^{-1}~deg^{-2}}$ in the
observed field with the galactic coordinate ($l$,
$b)=(62.\hspace{-.6ex}^\circ30$, $-29.\hspace{-.6ex}^\circ53$)
\citep{kow09}. The {\it Kepler} satellite has provided
statistical features of the flares of dwarf stars; the number of flares
with $t\geq20$~min is proportional to $t^{-2}$ \citep{dav14}. The expected
occurrence number of two flares somewhere in the field during the HSC
observing run is $\sim0.018$. The possibility is further reduced with a
condition that these flares take place 
at the position of the distant galaxy.
Therefore, a set of flares unlikely reproduce \SNa.

\section{Comparisons with theoretical models}
\label{sec:theoretical}

Since there are no known transients or variables that are consistent with the peak luminosity,
decline rate, and color of \SNa, we compare these properties of \SNa\ with published
theoretical models suggested for objects with a rapid rise and/or decline; (1)
shock breakout at the stellar surface of SN~IIP \citep{tom11}, (2) 
emission by the cooling envelopes of SN~IIb \citep{tsv12}, and (3) the explosion
of red supergiant star with a dense circumstellar wind \citep{moriya18}. All models
are calculated with the multigroup radiation hydrodynamics code
{\sc STELLA} \citep{bli06} and the synthetic spectral energy
distribution is convolved with the CCD quantum efficiency, transmittance
of the dewar window and the Primary Focus Unit, and filter transmission
curves of HSC.\footnote{{\url
http://www.naoj.org/Observing/Instruments/HSC/sensitivity.html}}

\begin{figure*}
\plotone{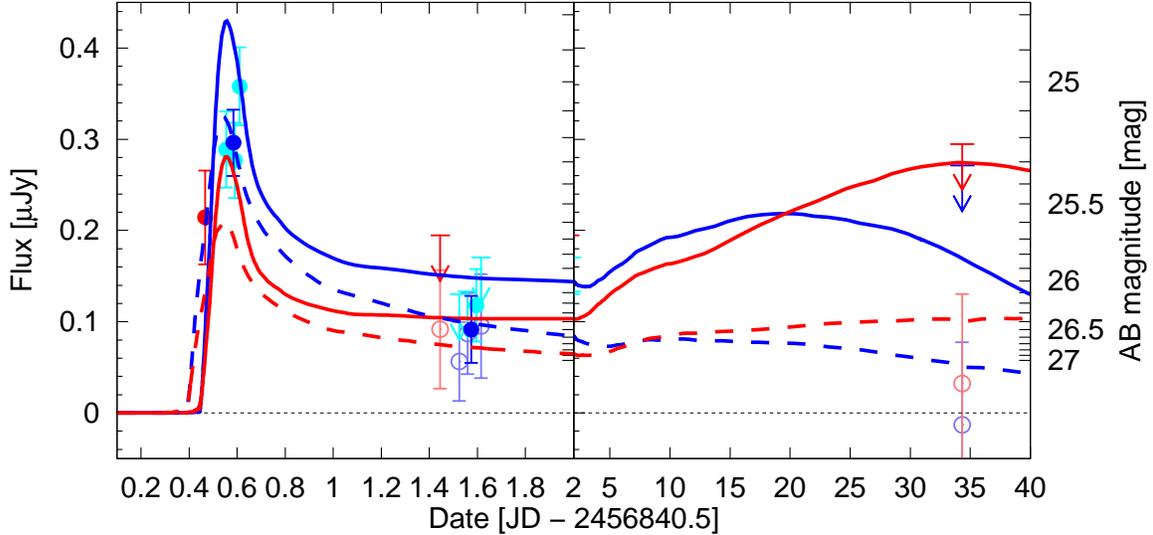}
\caption{Comparisons between the $g$-band (blue) and $r$-band (red)
 light curves of \SNa\ (symbols, same as Figure~\ref{fig:lc}) and the
 shock breakout at the stellar surface of the low-energy explosions with
 $E=0.15\times10^{51}$~erg (dashed lines)
 and $E=0.4\times10^{51}$~ergs (solid lines). 
}
\label{fig:modelSBO}
\end{figure*}

Figure~\ref{fig:model} shows comparisons between \SNa\ and the
theoretical models at epochs with absolute magnitudes similar to that of
\SNa\ ($\Mg\sim-16.2$~mag) and with similar $\Rgd$. The high decline rate with $\Mg\sim-16.2$~mag is
realized in the decline after the shock breakout peak and the cooling
envelope peak immediately after the explosion for mechanisms (1) and (2),
respectively, while it is achieved in the decline immediately after the
forward shock breaks out from a dense wind with a high mass loss rate
of $10^{-2}~\Msun~{\rm yr^{-1}}$ for mechanism (3) (top panels
of Figure~\ref{fig:model}). 

 \begin{figure}
  \epsscale{0.8}
\plotone{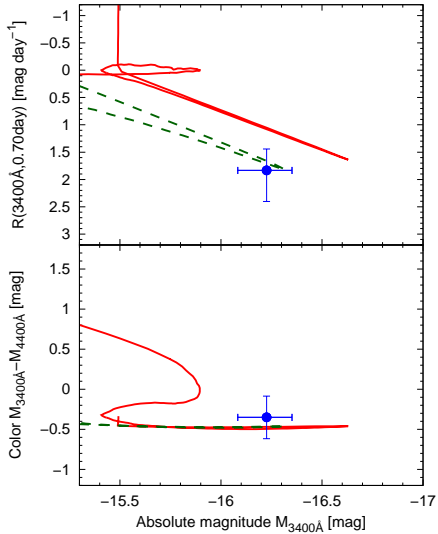}
\caption{Rate of brightness change (top) and color (bottom) of \SNa\ (blue) compared with those of the
 shock breakout at the stellar surface of the low-energy explosions with
 $E=0.15\times10^{51}$~erg (green)
 and $E=0.4\times10^{51}$~ergs (red). 
}
\label{fig:modelSBO2}
 \end{figure}

Middle panels of Figure~\ref{fig:model} show $\Rgd$ as a function of
$\Mg$. The rates of brightness change with $\Mg\sim-16.2$~mag are as high as $\sim+1~\Munit$ for
mechanisms (1) and (2) and $\sim+1.4~\Munit$ for mechanism (3). These are
all slightly slower than that of \SNa\. The bottom panels of
Figure~\ref{fig:model} show $\Mg-\Mr$ as a function of $\Mg$. The color
and absolute magnitude of \SNa\ is reproduced with mechanisms (1) and
(2), while the color of mechanism (3) is $0.8$~mag redder than \SNa.
The large color difference indicates that mechanism (3) is not the origin of \SNa.

The middle panel of Figure~\ref{fig:model} demonstrates that the decline
rate of mechanism (2) is independent on the explosion energy and the
presupernova radius. The rate of brightness change of $\Rgd\sim+1~\Munit$
is consistent with those of SNe~IIb observed to date
(section~\ref{sec:otherSN}), and slower than that of \SNa, 
although the decline rate could be higher if relativistic corrections are taken
into account \citep{tol13}. The influence of relativistic corrections 
requires more detailed radiative transfer modeling from optically thin
(transparent) to optically thick (diffusion) regions and should be studied
elsewhere.

On the other hand, a rate of brightness change as high as $\Rgd\sim+1.5~\Munit$ 
appears in explosions of stars with $\Mms=25$ and $30\Msun$,
corresponding to stellar radii of $1200$ and $1360\Rsun$, in mechanism (1). 
The top left panel of Figure~\ref{fig:model} demonstrates that the slow
decline rate stems from the rapid rising after the decline. As the peak
magnitude and the rising after the decline depend on the explosion
energy, we additionally calculate properties
for $25\Msun$ models with lower
explosion energies. Figure~\ref{fig:modelSBO} and
Figure~\ref{fig:modelSBO2} demonstrate that the peak magnitude, the
decline rate, and the color of \SNa\ are reproduced by mechanism (1)
with lower explosion energies of
$E\leq0.4\times10^{51}$~ergs. The low-energy explosions
are also consistent with the non-detection of \SNa\ on Day~35. The
absolute plateau brightness of the low-energy explosions is located at
the fainter side of the luminosity function of
nearby SNe~IIP \citep{ric14}. Thus, \SNa\ is most likely to be a shock breakout
at the stellar surface of a low-energy SN~IIP explosion.

\section{DISCUSSION \& CONCLUSIONS}
\label{sec:conclusion} 

We perform a high-cadence survey with 
Subaru/HSC and carry out follow-up observations as part of SHOOT. In this
paper, we focus on \SNa\, which rapidly declines in 
observations over two successive nights.

Spectroscopic follow-up observations of the host galaxy reveal that \SNa\ takes
place at $z=0.42285$. Based on the photometric observations, we
examine the nature of \SNa. \SNa\ is unlikely to be an active galactic
nucleus or a tidal disruption event because it appeared at the outskirt of
the host galaxy. The distance of \SNa\ from the center of the host
galaxy ($2.5$~kpc) is comparable with those of \SNsnls\ ($\sim1.7$~kpc) and
PS1-13arp ($4.5$~kpc). The peak
luminosity, decline rate, and color of \SNa\ are inconsistent with those of 
SNe at $>1$~day after the explosions, rapid transients, AT2017gfo,
AT2018cow, GRB prompt emission, or a GRB orphan
afterglow. Some properties of \SNa\ are consistent with those of some
transients. For example, the decline of AT2017gfo is as fast as that of
\SNa\, and the luminosity and decline rate of \SNa\ can be reproduced by
a GRB orphan afterglow. However, AT2017gfo is fainter and redder than
\SNa, and a GRB orphan afterglow is redder than \SNa. Furthermore, their 
probability of appearing in our survey expected is much smaller than unity
(Section~\ref{sec:otherOther} and \citealt{sco18}).
A set of flares of
a low-mass star, which could reproduce the light curve of \SNa, 
are unlikely to coincide with a distant galaxy.

Since the observational properties of \SNa\ are
inconsistent with the known transients or variables, we compared the properties
against published theoretical models calculated with {\sc STELLA}; (1) 
shock breakout at the stellar surface of SN~IIP, (2) emission by 
the cooling envelope of SN~IIb, and (3) the explosion of a red supergiant 
star with a dense circumstellar wind. 
None of them is fully consistent with \SNa.
They have slightly slower rates of brightness change than \SNa\ at $\Mg\sim-16.2$~mag.
Furthermore, while the color of \SNa\ is consistent 
with mechanisms (1) and (2), mechanism (3) gives a color redder by $+0.8$~mag 
than \SNa.

The inconsistency between \SNa\ and mechanism (1) can be solved
with low-energy explosions of a star with $25\Msun$. The multicolor
light curves and color of \SNa\ are reproduced by mechanism (1) with the
SN~IIP explosion of a progenitor star with $\Rs=1200\Rsun$ and
$E\leq0.4\times10^{51}$~ergs. The low-energy
explosion model is consistent with the non-detection of \SNa\ on Day~35.
Thus, we suggest that \SNa\ is the shock breakout at the stellar surface
of a low-energy SN~IIP explosion. 

An event rate of \SNa\ is roughly estimated by using a $1/V_{\rm max}$
method \citep{sch68,eal93} as done in \cite{tanaka16}. The event rate $R$ is written as
$R=1/p\tau \Omega V_{\rm max}$, where $p$ is a detection efficiency, $\tau$ is the
longer of a rest-frame time scale of a transient $\tau_{\rm tran}$ and a
survey duration of each field $\tau_{\rm obs,field}$, $\Omega$ is a field of view of HSC, and
$V_{\rm max}$ is a maximum volume per width in which a transient can be detected. Here, we assume $p=1$ because it is difficult to
evaluate $p$ for our complicated selection criteria. The observed
duration of each field in the $g$-band and in each night varies from
$1$~hr to $5.7$~hr, and thus $\tau V_{\rm max}$ can be derived from
\[
 \tau V_{\rm max} = {1\over{4\pi}}\sum_{\rm field} \int_0^{z_{\rm max}} \max\left\{\tau_{\rm tran},{\tau_{\rm obs,field}\over{1+z}}\right\}
 {dV\over{dz}} dz,
\]
where $z_{\rm max}$ is the maximum redshift in which a transient can be
detected and ${dV\over{dz}}$ is the comoving volume element per unit redshift.
Based on the light curve of \SNa, we adopt
the following two cases of $\tau$ and absolute magnitude $M$: (1) $\tau_{\rm tran}=0.1$~days and
$M=-16.2$~mag and (2) $\tau_{\rm tran}=0.7$~days and $M=-14.9$~mag.
Adopting the $3\sigma$ limiting magnitude of $26.3$~mag in the $g$-band,
the resultant $\tau V_{\rm max}$ are
$\tau V_{\rm max}=2.2\times10^{-3}$~days~Gpc$^3$~deg$^{-2}$ for case (1) and
$1.9\times10^{-3}$~days~Gpc$^3$~deg$^{-2}$ for case (2). Therefore, the
event rates are $9.4\times10^{-5}$~yr$^{-1}$~Mpc$^{-3}$ for case (1) and
$1.1\times10^{-4}$~yr$^{-1}$~Mpc$^{-3}$ for case (2). These are slightly
lower than the core-collapse SN rate of
$(3-7)\times10^{-4}$~yr$^{-1}$~Mpc$^{-3}$ at $z=0-1$
\citep{dah04,dah12,bot08,li11part2}.

According to \cite{for18}, taking
into account the bias toward the high detectability of SNe~IIP with
dense CSM in their observation, one fifth of SNe~IIP may exhibit 
shock breakout at the stellar surface. The rate of \SNa\ is
consistent with the suggestion that \SNa\ is the shock breakout 
at the stellar surface
of a SNe~IIP. This indicates that the rapidly declining transient as \SNa\
is a proxy for a massive star without dense CSM. The rapidly
rising/declining transients are representative of massive stars with
various CSM and their rates can constrain the fraction of massive stars
with/without dense CSM. While rapidly rising transients have
received a lot of attention recently, 
rapidly declining transients are also important
for understanding the fate of massive stars. Although the detection of
rapidly declining transients is more difficult than that of 
rapidly rising transients, it is important to consider a strategy to
detect rapidly declining transients.

The absolute magnitude of host galaxy is $M\sim-17$~mag. This is located
at the fainter end of the host galaxies of core-collapse supernovae in
the local Universe \citep{arc10}. The stellar mass and specific star
formation rate of the host galaxy is located at the smaller and higher ends
of host galaxies of core-collapse supernovae in the local Universe,
respectively, and the metallicity of the host galaxy is lower
than the typical value of host galaxies of SNe IIP
\citep[\eg][]{gra17}. SNe~IIP in low-mass host galaxies
with low metallicity are actually found in the local Universe \citep[\eg][]{gut18}. While
the star formation rate of the host galaxy of \SNa\ is smaller than
those of host galaxies of high-$z$ core-collapse supernovae observed so
far, its stellar mass and specific star formation rate 
are located at
the lower ends of these distributions \citep{sve10}. 
These properties of the host galaxy
indicate that a core-collapse supernova can occur in it.

If \SNa\ is the shock breakout at the stellar surface, the main-sequence mass of the best-fit
model is larger than constraints obtained for the masses of
progenitors of nearby
SNe~IIP \citep{sma09b,des13,gon15}. However, these observations are
limited to the local Universe because of the requirements of
presupernova images and late nebular spectra. The main-sequence
masses of SNe~IIP have never been measured at a redshift as high as
$z=0.4$. The low-metallicity environment may allow a massive red supergiant star
to survive until the SN explosion. 
Although it needs a careful treatment, due to a possible
observational bias that the shock breakout is more easily detected for
larger $\Mms$ because of larger $\Rs$, high-cadence observations
might provide a new clue to investigate the presupernova radius and $\Mms$
of SN progenitors
outside the local Universe.

The discovery of \SNa\ and other rapidly rising transients in
\cite{tanaka16} demonstrates that a high-cadence multicolor optical
transient survey with intervals of about one hour is sensitive to
high-$z$ transients. Unfortunately, the nature of \SNa\ and other
rapidly-rising transients is not well constrained due to the lack of 
immediate and continuous follow-up observations. Extensive high-cadence
multicolor surveys together with immediate and continuous follow-up
observations will provide the clue to investigate the high-$z$ Universe
with short-timescale transients.

\acknowledgments

N.T. thanks Suvi Gezari for clarifying the data quality of PS1-13arp,
Yuki~Kikuchi, Yuki~Taniguchi, and Takahiro~Kato for helping the
observations, and Michael W. Richmond for improving the English grammar
of the manuscript.
Data analyses were in part carried out on PC cluster
at Center for Computational Astrophysics, National Astronomical
Observatory of Japan. 
This research has been supported in part by the research grant program
of Toyota foundation (D11-R-0830), the
RFBR (19-52-50014)-JSPS bilateral program, World Premier
International Research Center Initiative, MEXT, Japan, and by the
Grant-in-Aid for Scientific Research of the JSPS (JP23224004, JP23540262,
JP23740157, JP15H05440, JP15H02075, JP16H02166, JP17H06363)
and MEXT (JP23105705). 
TJM is supported by Japan Society for the Promotion of Science
Postdoctoral Fellowships for Research Abroad (26\textperiodcentered 51).
This work of S.Blinnikov  (hydrodynamics of shock breakout and
development  of STELLA code) was supported by Russian Science
Foundation grant 19-12-00229.
This paper makes use of software developed for the Large Synoptic Survey
Telescope. We thank the LSST Project for making their code available as
free software at {\url http://dm.lsstcorp.org}.
Funding for SDSS-III has been provided by the Alfred P. Sloan
Foundation, the Participating Institutions, the National Science
Foundation, and the U.S. Department of Energy Office of Science. The
SDSS-III web site is {\url http://www.sdss3.org/}.
SDSS-III is managed by the Astrophysical Research Consortium for the
Participating Institutions of the SDSS-III Collaboration including the
University of Arizona, the Brazilian Participation Group, Brookhaven
National Laboratory, Carnegie Mellon University, University of Florida,
the French Participation Group, the German Participation Group, Harvard
University, the Instituto de Astrofisica de Canarias, the Michigan
State/Notre Dame/JINA Participation Group, Johns Hopkins University,
Lawrence Berkeley National Laboratory, Max Planck Institute for
Astrophysics, Max Planck Institute for Extraterrestrial Physics, New
Mexico State University, New York University, Ohio State University,
Pennsylvania State University, University of Portsmouth, Princeton
University, the Spanish Participation Group, University of Tokyo,
University of Utah, Vanderbilt University, University of Virginia,
University of Washington, and Yale University. 

\facilities{Subaru (HSC, FOCAS), Gemini-N (GMOS-N)}

\software{STELLA \citep{bli06}, IRAF \cite{tod86,tod93},
Gemini-IRAF\footnote{{\url https://www.gemini.edu/node/11823}}, HSC pipeline (v3.6.1; \citealt{bos18}), LSST pipeline
\citep{ive08,axe10}, HOTPANTS \citep{bec15}, WCSTools \citep{min02},
Astropy \citep{astropy18}, SExtractor \citep{ber96}}

\bibliographystyle{apj} 
\bibliography{ms}

\begin{thebibliography}{}
\expandafter\ifx\csname natexlab\endcsname\relax\def\natexlab#1{#1}\fi

\bibitem[{{Abbott} {et~al.}(2017){Abbott}, {Abbott}, {Abbott}, {Acernese},
  {Ackley}, {Adams}, {Adams}, {Addesso}, {Adhikari}, {Adya}, \&
  et~al.}]{GW170817MMApaper}
{Abbott}, B.~P., {Abbott}, R., {Abbott}, T.~D., {et~al.} 2017, \apjl, 848, L12

\bibitem[{{Aihara} {et~al.}(2011){Aihara}, {Allende Prieto}, {An}, {Anderson},
  {Aubourg}, {Balbinot}, {Beers}, {Berlind}, {Bickerton}, {Bizyaev}, {Blanton},
  {Bochanski}, {Bolton}, {Bovy}, {Brandt}, {Brinkmann}, {Brown}, {Brownstein},
  {Busca}, {Campbell}, {Carr}, {Chen}, {Chiappini}, {Comparat}, {Connolly},
  {Cortes}, {Croft}, {Cuesta}, {da Costa}, {Davenport}, {Dawson}, {Dhital},
  {Ealet}, {Ebelke}, {Edmondson}, {Eisenstein}, {Escoffier}, {Esposito},
  {Evans}, {Fan}, {Femen{\'{\i}}a Castell{\'a}}, {Font-Ribera}, {Frinchaboy},
  {Ge}, {Gillespie}, {Gilmore}, {Gonz{\'a}lez Hern{\'a}ndez}, {Gott}, {Gould},
  {Grebel}, {Gunn}, {Hamilton}, {Harding}, {Harris}, {Hawley}, {Hearty}, {Ho},
  {Hogg}, {Holtzman}, {Honscheid}, {Inada}, {Ivans}, {Jiang}, {Johnson},
  {Jordan}, {Jordan}, {Kazin}, {Kirkby}, {Klaene}, {Knapp}, {Kneib},
  {Kochanek}, {Koesterke}, {Kollmeier}, {Kron}, {Lampeitl}, {Lang}, {Le Goff},
  {Lee}, {Lin}, {Long}, {Loomis}, {Lucatello}, {Lundgren}, {Lupton}, {Ma},
  {MacDonald}, {Mahadevan}, {Maia}, {Makler}, {Malanushenko}, {Malanushenko},
  {Mandelbaum}, {Maraston}, {Margala}, {Masters}, {McBride}, {McGehee},
  {McGreer}, {M{\'e}nard}, {Miralda-Escud{\'e}}, {Morrison}, {Mullally},
  {Muna}, {Munn}, {Murayama}, {Myers}, {Naugle}, {Neto}, {Nguyen}, {Nichol},
  {O'Connell}, {Ogando}, {Olmstead}, {Oravetz}, {Padmanabhan},
  {Palanque-Delabrouille}, {Pan}, {Pandey}, {P{\^a}ris}, {Percival},
  {Petitjean}, {Pfaffenberger}, {Pforr}, {Phleps}, {Pichon}, {Pieri}, {Prada},
  {Price-Whelan}, {Raddick}, {Ramos}, {Reyl{\'e}}, {Rich}, {Richards}, {Rix},
  {Robin}, {Rocha-Pinto}, {Rockosi}, {Roe}, {Rollinde}, {Ross}, {Ross},
  {Rossetto}, {S{\'a}nchez}, {Sayres}, {Schlegel}, {Schlesinger}, {Schmidt},
  {Schneider}, {Sheldon}, {Shu}, {Simmerer}, {Simmons}, {Sivarani}, {Snedden},
  {Sobeck}, {Steinmetz}, {Strauss}, {Szalay}, {Tanaka}, {Thakar}, {Thomas},
  {Tinker}, {Tofflemire}, {Tojeiro}, {Tremonti}, {Vandenberg}, {Vargas
  Maga{\~n}a}, {Verde}, {Vogt}, {Wake}, {Wang}, {Weaver}, {Weinberg}, {White},
  {White}, {Yanny}, {Yasuda}, {Yeche}, \& {Zehavi}}]{aih11}
{Aihara}, H., {Allende Prieto}, C., {An}, D., {et~al.} 2011, \apjs, 193, 29

\bibitem[{{Alard}(2000)}]{ala00}
{Alard}, C. 2000, \aaps, 144, 363

\bibitem[{{Alard} \& {Lupton}(1998)}]{ala98}
{Alard}, C., \& {Lupton}, R.~H. 1998, \apj, 503, 325

\bibitem[{{Andreoni} {et~al.}(2017){Andreoni}, {Ackley}, {Cooke}, {Acharyya},
  {Allison}, {Anderson}, {Ashley}, {Baade}, {Bailes}, {Bannister}, {Beardsley},
  {Bessell}, {Bian}, {Bland}, {Boer}, {Booler}, {Brandeker}, {Brown},
  {Buckley}, {Chang}, {Coward}, {Crawford}, {Crisp}, {Crosse}, {Cucchiara},
  {Cup{\'a}k}, {de Gois}, {Deller}, {Devillepoix}, {Dobie}, {Elmer}, {Emrich},
  {Farah}, {Farrell}, {Franzen}, {Gaensler}, {Galloway}, {Gendre}, {Giblin},
  {Goobar}, {Green}, {Hancock}, {Hartig}, {Howell}, {Horsley}, {Hotan},
  {Howie}, {Hu}, {Hu}, {James}, {Johnston}, {Johnston-Hollitt}, {Kaplan},
  {Kasliwal}, {Keane}, {Kenney}, {Klotz}, {Lau}, {Laugier}, {Lenc}, {Li},
  {Liang}, {Lidman}, {Luvaul}, {Lynch}, {Ma}, {Macpherson}, {Mao},
  {McClelland}, {McCully}, {M{\"o}ller}, {Morales}, {Morris}, {Murphy},
  {Noysena}, {Onken}, {Orange}, {Os{\l}owski}, {Pallot}, {Paxman}, {Potter},
  {Pritchard}, {Raja}, {Ridden-Harper}, {Romero-Colmenero}, {Sadler}, {Sansom},
  {Scalzo}, {Schmidt}, {Scott}, {Seghouani}, {Shang}, {Shannon}, {Shao},
  {Shara}, {Sharp}, {Sokolowski}, {Sollerman}, {Staff}, {Steele}, {Sun},
  {Suntzeff}, {Tao}, {Tingay}, {Towner}, {Thierry}, {Trott}, {Tucker},
  {V{\"a}is{\"a}nen}, {Krishnan}, {Walker}, {Wang}, {Wang}, {Wayth}, {Whiting},
  {Williams}, {Williams}, {Wolf}, {Wu}, {Wu}, {Yang}, {Yuan}, {Zhang}, {Zhou},
  \& {Zovaro}}]{GW170817andreoni}
{Andreoni}, I., {Ackley}, K., {Cooke}, J., {et~al.} 2017, \pasa, 34, e069

\bibitem[{{Arcavi} {et~al.}(2010){Arcavi}, {Gal-Yam}, {Kasliwal}, {Quimby},
  {Ofek}, {Kulkarni}, {Nugent}, {Cenko}, {Bloom}, {Sullivan}, {Howell},
  {Poznanski}, {Filippenko}, {Law}, {Hook}, {J{\"o}nsson}, {Blake}, {Cooke},
  {Dekany}, {Rahmer}, {Hale}, {Smith}, {Zolkower}, {Velur}, {Walters},
  {Henning}, {Bui}, {McKenna}, \& {Jacobsen}}]{arc10}
{Arcavi}, I., {Gal-Yam}, A., {Kasliwal}, M.~M., {et~al.} 2010, \apj, 721, 777

\bibitem[{{Arcavi} {et~al.}(2017){Arcavi}, {Hosseinzadeh}, {Howell}, {McCully},
  {Poznanski}, {Kasen}, {Barnes}, {Zaltzman}, {Vasylyev}, {Maoz}, \&
  {Valenti}}]{GW170817arcavi}
{Arcavi}, I., {Hosseinzadeh}, G., {Howell}, D.~A., {et~al.} 2017, \nat, 551, 64

\bibitem[{{Astropy Collaboration} {et~al.}(2018){Astropy Collaboration},
  {Price-Whelan}, {Sip{\H{o}}cz}, {G{\"u}nther}, {Lim}, {Crawford}, {Conseil},
  {Shupe}, {Craig}, {Dencheva}, {Ginsburg}, {Vand erPlas}, {Bradley},
  {P{\'e}rez-Su{\'a}rez}, {de Val-Borro}, {Aldcroft}, {Cruz}, {Robitaille},
  {Tollerud}, {Ardelean}, {Babej}, {Bach}, {Bachetti}, {Bakanov}, {Bamford},
  {Barentsen}, {Barmby}, {Baumbach}, {Berry}, {Biscani}, {Boquien}, {Bostroem},
  {Bouma}, {Brammer}, {Bray}, {Breytenbach}, {Buddelmeijer}, {Burke},
  {Calderone}, {Cano Rodr{\'\i}guez}, {Cara}, {Cardoso}, {Cheedella}, {Copin},
  {Corrales}, {Crichton}, {D'Avella}, {Deil}, {Depagne}, {Dietrich}, {Donath},
  {Droettboom}, {Earl}, {Erben}, {Fabbro}, {Ferreira}, {Finethy}, {Fox},
  {Garrison}, {Gibbons}, {Goldstein}, {Gommers}, {Greco}, {Greenfield},
  {Groener}, {Grollier}, {Hagen}, {Hirst}, {Homeier}, {Horton}, {Hosseinzadeh},
  {Hu}, {Hunkeler}, {Ivezi{\'c}}, {Jain}, {Jenness}, {Kanarek}, {Kendrew},
  {Kern}, {Kerzendorf}, {Khvalko}, {King}, {Kirkby}, {Kulkarni}, {Kumar},
  {Lee}, {Lenz}, {Littlefair}, {Ma}, {Macleod}, {Mastropietro}, {McCully},
  {Montagnac}, {Morris}, {Mueller}, {Mumford}, {Muna}, {Murphy}, {Nelson},
  {Nguyen}, {Ninan}, {N{\"o}the}, {Ogaz}, {Oh}, {Parejko}, {Parley}, {Pascual},
  {Patil}, {Patil}, {Plunkett}, {Prochaska}, {Rastogi}, {Reddy Janga},
  {Sabater}, {Sakurikar}, {Seifert}, {Sherbert}, {Sherwood-Taylor}, {Shih},
  {Sick}, {Silbiger}, {Singanamalla}, {Singer}, {Sladen}, {Sooley},
  {Sornarajah}, {Streicher}, {Teuben}, {Thomas}, {Tremblay}, {Turner},
  {Terr{\'o}n}, {van Kerkwijk}, {de la Vega}, {Watkins}, {Weaver}, {Whitmore},
  {Woillez}, {Zabalza}, \& {Astropy Contributors}}]{astropy18}
{Astropy Collaboration}, {Price-Whelan}, A.~M., {Sip{\H{o}}cz}, B.~M., {et~al.}
  2018, \aj, 156, 123

\bibitem[{{Axelrod} {et~al.}(2010){Axelrod}, {Kantor}, {Lupton}, \&
  {Pierfederici}}]{axe10}
{Axelrod}, T., {Kantor}, J., {Lupton}, R.~H., \& {Pierfederici}, F. 2010, in
  Society of Photo-Optical Instrumentation Engineers (SPIE) Conference Series,
  Vol. 7740, Society of Photo-Optical Instrumentation Engineers (SPIE)
  Conference Series, 15

\bibitem[{{Becker}(2015)}]{bec15}
{Becker}, A. 2015, {HOTPANTS: High Order Transform of PSF ANd Template
  Subtraction}, Astrophysics Source Code Library, ascl:1504.004

\bibitem[{{Bersten} {et~al.}(2018){Bersten}, {Folatelli}, {Garc{\'{\i}}a}, {van
  Dyk}, {Benvenuto}, {Orellana}, {Buso}, {S{\'a}nchez}, {Tanaka}, {Maeda},
  {Filippenko}, {Zheng}, {Brink}, {Cenko}, {de Jaeger}, {Kumar}, {Moriya},
  {Nomoto}, {Perley}, {Shivvers}, \& {Smith}}]{ber18}
{Bersten}, M.~C., {Folatelli}, G., {Garc{\'{\i}}a}, F., {et~al.} 2018, \nat,
  554, 497

\bibitem[{{Bertin} \& {Arnouts}(1996)}]{ber96}
{Bertin}, E., \& {Arnouts}, S. 1996, \aaps, 117, 393

\bibitem[{{Blinnikov} {et~al.}(2006){Blinnikov}, {R{\"o}pke}, {Sorokina},
  {Gieseler}, {Reinecke}, {Travaglio}, {Hillebrandt}, \& {Stritzinger}}]{bli06}
{Blinnikov}, S.~I., {R{\"o}pke}, F.~K., {Sorokina}, E.~I., {et~al.} 2006, \aap,
  453, 229

\bibitem[{{Bosch} {et~al.}(2018){Bosch}, {Armstrong}, {Bickerton}, {Furusawa},
  {Ikeda}, {Koike}, {Lupton}, {Mineo}, {Price}, {Takata}, {Tanaka}, {Yasuda},
  {AlSayyad}, {Becker}, {Coulton}, {Coupon}, {Garmilla}, {Huang}, {Krughoff},
  {Lang}, {Leauthaud}, {Lim}, {Lust}, {MacArthur}, {Mandelbaum}, {Miyatake},
  {Miyazaki}, {Murata}, {More}, {Okura}, {Owen}, {Swinbank}, {Strauss},
  {Yamada}, \& {Yamanoi}}]{bos18}
{Bosch}, J., {Armstrong}, R., {Bickerton}, S., {et~al.} 2018, \pasj, 70, S5

\bibitem[{{Botticella} {et~al.}(2008){Botticella}, {Riello}, {Cappellaro},
  {Benetti}, {Altavilla}, {Pastorello}, {Turatto}, {Greggio}, {Patat},
  {Valenti}, {Zampieri}, {Harutyunyan}, {Pignata}, \& {Taubenberger}}]{bot08}
{Botticella}, M.~T., {Riello}, M., {Cappellaro}, E., {et~al.} 2008, \aap, 479,
  49

\bibitem[{{Brown} {et~al.}(2012){Brown}, {Dawson}, {de Pasquale}, {Gronwall},
  {Holland}, {Immler}, {Kuin}, {Mazzali}, {Milne}, {Oates}, \&
  {Siegel}}]{bro12}
{Brown}, P.~J., {Dawson}, K.~S., {de Pasquale}, M., {et~al.} 2012, \apj, 753,
  22

\bibitem[{{Bruzual} \& {Charlot}(2003)}]{bur03}
{Bruzual}, G., \& {Charlot}, S. 2003, \mnras, 344, 1000

\bibitem[{{Chabrier}(2003)}]{cha03}
{Chabrier}, G. 2003, \pasp, 115, 763

\bibitem[{{Chen} {et~al.}(2019){Chen}, {Dong}, {Stritzinger}, {Holmbo},
  {Strader}, {Kochanek}, {Peng}, {Benetti}, {Bersier}, \&
  {Brownsberger}}]{che19}
{Chen}, P., {Dong}, S., {Stritzinger}, M.~D., {et~al.} 2019, arXiv e-prints,
  arXiv:1905.02205

\bibitem[{{Chevalier} \& {Irwin}(2011)}]{che11}
{Chevalier}, R.~A., \& {Irwin}, C.~M. 2011, \apjl, 729, L6

\bibitem[{{Coulter} {et~al.}(2017){Coulter}, {Foley}, {Kilpatrick}, {Drout},
  {Piro}, {Shappee}, {Siebert}, {Simon}, {Ulloa}, {Kasen}, {Madore},
  {Murguia-Berthier}, {Pan}, {Prochaska}, {Ramirez-Ruiz}, {Rest}, \&
  {Rojas-Bravo}}]{GW170817coulter}
{Coulter}, D.~A., {Foley}, R.~J., {Kilpatrick}, C.~D., {et~al.} 2017, Science,
  358, 1556

\bibitem[{{Cowperthwaite} {et~al.}(2017){Cowperthwaite}, {Berger}, {Villar},
  {Metzger}, {Nicholl}, {Chornock}, {Blanchard}, {Fong}, {Margutti},
  {Soares-Santos}, {Alexander}, {Allam}, {Annis}, {Brout}, {Brown}, {Butler},
  {Chen}, {Diehl}, {Doctor}, {Drout}, {Eftekhari}, {Farr}, {Finley}, {Foley},
  {Frieman}, {Fryer}, {Garc{\'{\i}}a-Bellido}, {Gill}, {Guillochon}, {Herner},
  {Holz}, {Kasen}, {Kessler}, {Marriner}, {Matheson}, {Neilsen}, {Quataert},
  {Palmese}, {Rest}, {Sako}, {Scolnic}, {Smith}, {Tucker}, {Williams},
  {Balbinot}, {Carlin}, {Cook}, {Durret}, {Li}, {Lopes}, {Louren{\c c}o},
  {Marshall}, {Medina}, {Muir}, {Mu{\~n}oz}, {Sauseda}, {Schlegel}, {Secco},
  {Vivas}, {Wester}, {Zenteno}, {Zhang}, {Abbott}, {Banerji}, {Bechtol},
  {Benoit-L{\'e}vy}, {Bertin}, {Buckley-Geer}, {Burke}, {Capozzi}, {Carnero
  Rosell}, {Carrasco Kind}, {Castander}, {Crocce}, {Cunha}, {D'Andrea}, {da
  Costa}, {Davis}, {DePoy}, {Desai}, {Dietrich}, {Drlica-Wagner}, {Eifler},
  {Evrard}, {Fernandez}, {Flaugher}, {Fosalba}, {Gaztanaga}, {Gerdes},
  {Giannantonio}, {Goldstein}, {Gruen}, {Gruendl}, {Gutierrez}, {Honscheid},
  {Jain}, {James}, {Jeltema}, {Johnson}, {Johnson}, {Kent}, {Krause}, {Kron},
  {Kuehn}, {Nuropatkin}, {Lahav}, {Lima}, {Lin}, {Maia}, {March}, {Martini},
  {McMahon}, {Menanteau}, {Miller}, {Miquel}, {Mohr}, {Neilsen}, {Nichol},
  {Ogando}, {Plazas}, {Roe}, {Romer}, {Roodman}, {Rykoff}, {Sanchez},
  {Scarpine}, {Schindler}, {Schubnell}, {Sevilla-Noarbe}, {Smith}, {Smith},
  {Sobreira}, {Suchyta}, {Swanson}, {Tarle}, {Thomas}, {Thomas}, {Troxel},
  {Vikram}, {Walker}, {Wechsler}, {Weller}, {Yanny}, \&
  {Zuntz}}]{GW170817cowperthwaite}
{Cowperthwaite}, P.~S., {Berger}, E., {Villar}, V.~A., {et~al.} 2017, \apjl,
  848, L17

\bibitem[{{Dahlen} {et~al.}(2012){Dahlen}, {Strolger}, {Riess}, {Mattila},
  {Kankare}, \& {Mobasher}}]{dah12}
{Dahlen}, T., {Strolger}, L.-G., {Riess}, A.~G., {et~al.} 2012, \apj, 757, 70

\bibitem[{{Dahlen} {et~al.}(2004){Dahlen}, {Strolger}, {Riess}, {Mobasher},
  {Chary}, {Conselice}, {Ferguson}, {Fruchter}, {Giavalisco}, {Livio}, {Madau},
  {Panagia}, \& {Tonry}}]{dah04}
{Dahlen}, T., {Strolger}, L., {Riess}, A.~G., {et~al.} 2004, \apj, 613, 189

\bibitem[{{Davenport} {et~al.}(2014){Davenport}, {Hawley}, {Hebb},
  {Wisniewski}, {Kowalski}, {Johnson}, {Malatesta}, {Peraza}, {Keil},
  {Silverberg}, {Jansen}, {Scheffler}, {Berdis}, {Larsen}, \& {Hilton}}]{dav14}
{Davenport}, J.~R.~A., {Hawley}, S.~L., {Hebb}, L., {et~al.} 2014, \apj, 797,
  122

\bibitem[{{Dessart} {et~al.}(2013){Dessart}, {Hillier}, {Waldman}, \&
  {Livne}}]{des13}
{Dessart}, L., {Hillier}, D.~J., {Waldman}, R., \& {Livne}, E. 2013, \mnras,
  433, 1745

\bibitem[{{D{\'{\i}}az} {et~al.}(2017){D{\'{\i}}az}, {Macri}, {Garcia Lambas},
  {Mendes de Oliveira}, {Nilo Castell{\'o}n}, {Ribeiro}, {S{\'a}nchez},
  {Schoenell}, {Abramo}, {Akras}, {Alcaniz}, {Artola}, {Beroiz}, {Bonoli},
  {Cabral}, {Camuccio}, {Castillo}, {Chavushyan}, {Coelho}, {Colazo},
  {Costa-Duarte}, {Cuevas Larenas}, {DePoy}, {Dom{\'{\i}}nguez Romero},
  {Dultzin}, {Fern{\'a}ndez}, {Garc{\'{\i}}a}, {Girardini}, {Gon{\c c}alves},
  {Gon{\c c}alves}, {Gurovich}, {Jim{\'e}nez-Teja}, {Kanaan}, {Lares}, {Lopes
  de Oliveira}, {L{\'o}pez-Cruz}, {Marshall}, {Melia}, {Molino}, {Padilla},
  {Pe{\~n}uela}, {Placco}, {Qui{\~n}ones}, {Ram{\'{\i}}rez Rivera}, {Renzi},
  {Riguccini}, {R{\'{\i}}os-L{\'o}pez}, {Rodriguez}, {Sampedro}, {Schneiter},
  {Sodr{\'e}}, {Starck}, {Torres-Flores}, {Tornatore}, \&
  {Zadro{\.z}ny}}]{GW170817diaz}
{D{\'{\i}}az}, M.~C., {Macri}, L.~M., {Garcia Lambas}, D., {et~al.} 2017,
  \apjl, 848, L29

\bibitem[{{Drake} {et~al.}(2009){Drake}, {Djorgovski}, {Mahabal}, {Beshore},
  {Larson}, {Graham}, {Williams}, {Christensen}, {Catelan}, {Boattini},
  {Gibbs}, {Hill}, \& {Kowalski}}]{dra09}
{Drake}, A.~J., {Djorgovski}, S.~G., {Mahabal}, A., {et~al.} 2009, \apj, 696,
  870

\bibitem[{{Drout} {et~al.}(2014){Drout}, {Chornock}, {Soderberg}, {Sanders},
  {McKinnon}, {Rest}, {Foley}, {Milisavljevic}, {Margutti}, {Berger},
  {Calkins}, {Fong}, {Gezari}, {Huber}, {Kankare}, {Kirshner}, {Leibler},
  {Lunnan}, {Mattila}, {Marion}, {Narayan}, {Riess}, {Roth}, {Scolnic},
  {Smartt}, {Tonry}, {Burgett}, {Chambers}, {Hodapp}, {Jedicke}, {Kaiser},
  {Magnier}, {Metcalfe}, {Morgan}, {Price}, \& {Waters}}]{dro14}
{Drout}, M.~R., {Chornock}, R., {Soderberg}, A.~M., {et~al.} 2014, \apj, 794,
  23

\bibitem[{{Drout} {et~al.}(2017){Drout}, {Piro}, {Shappee}, {Kilpatrick},
  {Simon}, {Contreras}, {Coulter}, {Foley}, {Siebert}, {Morrell}, {Boutsia},
  {Di Mille}, {Holoien}, {Kasen}, {Kollmeier}, {Madore}, {Monson},
  {Murguia-Berthier}, {Pan}, {Prochaska}, {Ramirez-Ruiz}, {Rest}, {Adams},
  {Alatalo}, {Ba{\~n}ados}, {Baughman}, {Beers}, {Bernstein}, {Bitsakis},
  {Campillay}, {Hansen}, {Higgs}, {Ji}, {Maravelias}, {Marshall}, {Moni Bidin},
  {Prieto}, {Rasmussen}, {Rojas-Bravo}, {Strom}, {Ulloa},
  {Vargas-Gonz{\'a}lez}, {Wan}, \& {Whitten}}]{GW170817drout}
{Drout}, M.~R., {Piro}, A.~L., {Shappee}, B.~J., {et~al.} 2017, Science, 358,
  1570

\bibitem[{{Eales}(1993)}]{eal93}
{Eales}, S. 1993, \apj, 404, 51

\bibitem[{{Evans} {et~al.}(2017){Evans}, {Cenko}, {Kennea}, {Emery}, {Kuin},
  {Korobkin}, {Wollaeger}, {Fryer}, {Madsen}, {Harrison}, {Xu}, {Nakar},
  {Hotokezaka}, {Lien}, {Campana}, {Oates}, {Troja}, {Breeveld}, {Marshall},
  {Barthelmy}, {Beardmore}, {Burrows}, {Cusumano}, {D'A{\`i}}, {D'Avanzo},
  {D'Elia}, {de Pasquale}, {Even}, {Fontes}, {Forster}, {Garcia}, {Giommi},
  {Grefenstette}, {Gronwall}, {Hartmann}, {Heida}, {Hungerford}, {Kasliwal},
  {Krimm}, {Levan}, {Malesani}, {Melandri}, {Miyasaka}, {Nousek}, {O'Brien},
  {Osborne}, {Pagani}, {Page}, {Palmer}, {Perri}, {Pike}, {Racusin}, {Rosswog},
  {Siegel}, {Sakamoto}, {Sbarufatti}, {Tagliaferri}, {Tanvir}, \&
  {Tohuvavohu}}]{GW170817evans}
{Evans}, P.~A., {Cenko}, S.~B., {Kennea}, J.~A., {et~al.} 2017, Science, 358,
  1565

\bibitem[{{F{\"o}rster} {et~al.}(2016){F{\"o}rster}, {Maureira}, {San
  Mart{\'\i}n}, {Hamuy}, {Mart{\'\i}nez}, {Huijse}, {Cabrera}, {Galbany}, {de
  Jaeger}, {Gonz{\'a}lez"ŸGait{\'a}n}, {Anderson}, {Kunkarayakti}, {Pignata},
  {Bufano}, {Litt{\'\i}n}, {Olivares}, {Medina}, {Smith}, {Vivas},
  {Est{\'e}vez}, {Mu{\~n}oz}, \& {Vera}}]{for16}
{F{\"o}rster}, F., {Maureira}, J.~C., {San Mart{\'\i}n}, J., {et~al.} 2016,
  \apj, 832, 155

\bibitem[{{F{\"o}rster} {et~al.}(2018){F{\"o}rster}, {Moriya}, {Maureira},
  {Anderson}, {Blinnikov}, {Bufano}, {Cabrera-Vives}, {Clocchiatti}, {de
  Jaeger}, {Est{\'e}vez}, {Galbany}, {Gonz{\'a}lez-Gait{\'a}n}, {Gr{\"a}fener},
  {Hamuy}, {Hsiao}, {Huentelemu}, {Huijse}, {Kuncarayakti}, {Mart{\'{\i}}nez},
  {Medina}, {Olivares E.}, {Pignata}, {Razza}, {Reyes}, {San Mart{\'{\i}}n},
  {Smith}, {Vera}, {Vivas}, {de Ugarte Postigo}, {Yoon}, {Ashall}, {Fraser},
  {Gal-Yam}, {Kankare}, {Le Guillou}, {Mazzali}, {Walton}, \& {Young}}]{for18}
{F{\"o}rster}, F., {Moriya}, T.~J., {Maureira}, J.~C., {et~al.} 2018, Nature
  Astronomy, doi:10.1038/s41550-018-0563-4

\bibitem[{{Furusawa} {et~al.}(2010){Furusawa}, {Yasuda}, {Okura}, {Nakata},
  {Mineo}, {Takata}, {Tanaka}, {Itoh}, {Katayama}, {Komiyama}, {Miyazaki},
  {Utsumi}, {Aihara}, {Uchida}, \& {Lupton}}]{fur10}
{Furusawa}, H., {Yasuda}, N., {Okura}, Y., {et~al.} 2010, in Society of
  Photo-Optical Instrumentation Engineers (SPIE) Conference Series, Vol. 7740,
  Society of Photo-Optical Instrumentation Engineers (SPIE) Conference Series,
  2

\bibitem[{{Furusawa} {et~al.}(2011){Furusawa}, {Okura}, {Mineo}, {Takata},
  {Nakata}, {Tanaka}, {Katayama}, {Itoh}, {Yasuda}, {Miyazaki}, {Komiyama},
  {Utsumi}, {Uchida}, \& {Aihara}}]{fur11}
{Furusawa}, H., {Okura}, Y., {Mineo}, S., {et~al.} 2011, \pasj, 63, 585

\bibitem[{{Furusawa} {et~al.}(2018){Furusawa}, {Koike}, {Takata}, {Okura},
  {Miyatake}, {Lupton}, {Bickerton}, {Price}, {Bosch}, {Yasuda}, {Mineo},
  {Yamada}, {Miyazaki}, {Nakata}, {Koshida}, {Komiyama}, {Utsumi},
  {Kawanomoto}, {Jeschke}, {Noumaru}, {Schubert}, {Iwata}, {Finet},
  {Fujiyoshi}, {Tajitsu}, {Terai}, \& {Lee}}]{fur18}
{Furusawa}, H., {Koike}, M., {Takata}, T., {et~al.} 2018, \pasj, 70, S3

\bibitem[{{Gal-Yam} {et~al.}(2014){Gal-Yam}, {Arcavi}, {Ofek}, {Ben-Ami},
  {Cenko}, {Kasliwal}, {Cao}, {Yaron}, {Tal}, {Silverman}, {Horesh}, {De Cia},
  {Taddia}, {Sollerman}, {Perley}, {Vreeswijk}, {Kulkarni}, {Nugent},
  {Filippenko}, \& {Wheeler}}]{gal14}
{Gal-Yam}, A., {Arcavi}, I., {Ofek}, E.~O., {et~al.} 2014, \nat, 509, 471

\bibitem[{{Garn} \& {Best}(2010)}]{gar10}
{Garn}, T., \& {Best}, P.~N. 2010, \mnras, 409, 421

\bibitem[{{Garnavich} {et~al.}(2016){Garnavich}, {Tucker}, {Rest}, {Shaya},
  {Olling}, {Kasen}, \& {Villar}}]{gar16}
{Garnavich}, P.~M., {Tucker}, B.~E., {Rest}, A., {et~al.} 2016, \apj, 820, 23

\bibitem[{{Gezari} {et~al.}(2008){Gezari}, {Dessart}, {Basa}, {Martin},
  {Neill}, {Woosley}, {Hillier}, {Bazin}, {Forster}, {Friedman}, {Le Du},
  {Mazure}, {Morrissey}, {Neff}, {Schiminovich}, \& {Wyder}}]{gez08}
{Gezari}, S., {Dessart}, L., {Basa}, S., {et~al.} 2008, \apjl, 683, L131

\bibitem[{{Gezari} {et~al.}(2015){Gezari}, {Jones}, {Sanders}, {Soderberg},
  {Hung}, {Heinis}, {Smartt}, {Rest}, {Scolnic}, {Chornock}, {Berger}, {Foley},
  {Huber}, {Price}, {Stubbs}, {Riess}, {Kirshner}, {Smith}, {Wood-Vasey},
  {Schiminovich}, {Martin}, {Burgett}, {Chambers}, {Flewelling}, {Kaiser},
  {Tonry}, \& {Wainscoat}}]{gez15}
{Gezari}, S., {Jones}, D.~O., {Sanders}, N.~E., {et~al.} 2015, \apj, 804, 28

\bibitem[{{Gonz{\'a}lez-Gait{\'a}n} {et~al.}(2015){Gonz{\'a}lez-Gait{\'a}n},
  {Tominaga}, {Molina}, {Galbany}, {Bufano}, {Anderson}, {Gutierrez},
  {F{\"o}rster}, {Pignata}, {Bersten}, {Howell}, {Sullivan}, {Carlberg}, {de
  Jaeger}, {Hamuy}, {Baklanov}, \& {Blinnikov}}]{gon15}
{Gonz{\'a}lez-Gait{\'a}n}, S., {Tominaga}, N., {Molina}, J., {et~al.} 2015,
  \mnras, 451, 2212

\bibitem[{{Graham} {et~al.}(2019){Graham}, {Kulkarni}, {Bellm}, {Adams},
  {Barbarino}, {Blagorodnova}, {Bodewits}, {Bolin}, {Brady}, {Cenko}, {Chang},
  {Coughlin}, {De}, {Eadie}, {Farnham}, {Feindt}, {Franckowiak}, {Fremling},
  {Gezari}, {Ghosh}, {Goldstein}, {Golkhou}, {Goobar}, {Ho}, {Huppenkothen},
  {Ivezi{\'c}}, {Jones}, {Juric}, {Kaplan}, {Kasliwal}, {Kelley}, {Kupfer},
  {Lee}, {Lin}, {Lunnan}, {Mahabal}, {Miller}, {Ngeow}, {Nugent}, {Ofek},
  {Prince}, {Rauch}, {van Roestel}, {Schulze}, {Singer}, {Sollerman}, {Taddia},
  {Yan}, {Ye}, {Yu}, {Barlow}, {Bauer}, {Beck}, {Belicki}, {Biswas}, {Brinnel},
  {Brooke}, {Bue}, {Bulla}, {Burruss}, {Connolly}, {Cromer}, {Cunningham},
  {Dekany}, {Delacroix}, {Desai}, {Duev}, {Feeney}, {Flynn}, {Frederick},
  {Gal-Yam}, {Giomi}, {Groom}, {Hacopians}, {Hale}, {Helou}, {Henning},
  {Hover}, {Hillenbrand}, {Howell}, {Hung}, {Imel}, {Ip}, {Jackson}, {Kaspi},
  {Kaye}, {Kowalski}, {Kramer}, {Kuhn}, {Landry}, {Laher}, {Mao}, {Masci},
  {Monkewitz}, {Murphy}, {Nordin}, {Patterson}, {Penprase}, {Porter},
  {Rebbapragada}, {Reiley}, {Riddle}, {Rigault}, {Rodriguez}, {Rusholme}, {van
  Santen}, {Shupe}, {Smith}, {Soumagnac}, {Stein}, {Surace}, {Szkody}, {Terek},
  {Van Sistine}, {van Velzen}, {Vestrand}, {Walters}, {Ward}, {Zhang}, \&
  {Zolkower}}]{gra19ZTF}
{Graham}, M.~J., {Kulkarni}, S.~R., {Bellm}, E.~C., {et~al.} 2019, \pasp, 131,
  078001

\bibitem[{{Granot} {et~al.}(2002){Granot}, {Panaitescu}, {Kumar}, \&
  {Woosley}}]{gra02}
{Granot}, J., {Panaitescu}, A., {Kumar}, P., \& {Woosley}, S.~E. 2002, \apjl,
  570, L61

\bibitem[{{Graur} {et~al.}(2017){Graur}, {Bianco}, {Huang}, {Modjaz},
  {Shivvers}, {Filippenko}, {Li}, \& {Eldridge}}]{gra17}
{Graur}, O., {Bianco}, F.~B., {Huang}, S., {et~al.} 2017, \apj, 837, 120

\bibitem[{{Guti{\'e}rrez} {et~al.}(2018){Guti{\'e}rrez}, {Anderson},
  {Sullivan}, {Dessart}, {Gonz{\'a}lez-Gaitan}, {Galbany}, {Dimitriadis},
  {Arcavi}, {Bufano}, {Chen}, {Dennefeld}, {Gromadzki}, {Haislip},
  {Hosseinzadeh}, {Howell}, {Inserra}, {Kankare}, {Leloudas}, {Maguire},
  {McCully}, {Morrell}, {Olivares E}, {Pignata}, {Reichart}, {Reynolds},
  {Smartt}, {Sollerman}, {Taddia}, {Tak{\'a}ts}, {Terreran}, {Valenti}, \&
  {Young}}]{gut18}
{Guti{\'e}rrez}, C.~P., {Anderson}, J.~P., {Sullivan}, M., {et~al.} 2018,
  \mnras, 479, 3232

\bibitem[{{Ho} {et~al.}(2019){Ho}, {Goldstein}, {Schulze}, {Khatami}, {Perley},
  {Ergon}, {Gal-Yam}, {Corsi}, {Andreoni}, {Barbarino}, {Bellm},
  {Blagorodnova}, {Bright}, {Burns}, {Cenko}, {Cunningham}, {De}, {Dekany},
  {Dugas}, {Fender}, {Fransson}, {Fremling}, {Goldstein}, {Graham}, {Hale},
  {Horesh}, {Hung}, {Kasliwal}, {Kuin}, {Kulkarni}, {Kupfer}, {Lunnan},
  {Masci}, {Ngeow}, {Nugent}, {Ofek}, {Patterson}, {Petitpas}, {Rusholme},
  {Sai}, {Sfaradi}, {Shupe}, {Sollerman}, {Soumagnac}, {Tachibana}, {Taddia},
  {Walters}, {Wang}, {Yao}, \& {Zhang}}]{ho19}
{Ho}, A. Y.~Q., {Goldstein}, D.~A., {Schulze}, S., {et~al.} 2019, arXiv
  e-prints, arXiv:1904.11009

\bibitem[{{Hook} {et~al.}(2004){Hook}, {J{\o}rgensen}, {Allington-Smith},
  {Davies}, {Metcalfe}, {Murowinski}, \& {Crampton}}]{hoo04}
{Hook}, I.~M., {J{\o}rgensen}, I., {Allington-Smith}, J.~R., {et~al.} 2004,
  \pasp, 116, 425

\bibitem[{{Ivezic} {et~al.}(2008){Ivezic}, {Tyson}, {Allsman}, {Andrew},
  {Angel}, \& {for the LSST Collaboration}}]{ive08}
{Ivezic}, Z., {Tyson}, J.~A., {Allsman}, R., {et~al.} 2008, ArXiv e-prints,
  arXiv:0805.2366

\bibitem[{{Juneau} {et~al.}(2014){Juneau}, {Bournaud}, {Charlot}, {Daddi},
  {Elbaz}, {Trump}, {Brinchmann}, {Dickinson}, {Duc}, {Gobat}, {Jean-Baptiste},
  {Le Floc'h}, {Lehnert}, {Pacifici}, {Pannella}, \& {Schreiber}}]{jun14}
{Juneau}, S., {Bournaud}, F., {Charlot}, S., {et~al.} 2014, \apj, 788, 88

\bibitem[{{Kashikawa} {et~al.}(2002){Kashikawa}, {Aoki}, {Asai}, {Ebizuka},
  {Inata}, {Iye}, {Kawabata}, {Kosugi}, {Ohyama}, {Okita}, {Ozawa}, {Saito},
  {Sasaki}, {Sekiguchi}, {Shimizu}, {Taguchi}, {Takata}, {Yadoumaru}, \&
  {Yoshida}}]{kas02}
{Kashikawa}, N., {Aoki}, K., {Asai}, R., {et~al.} 2002, \pasj, 54, 819

\bibitem[{{Kasliwal} {et~al.}(2017){Kasliwal}, {Nakar}, {Singer}, {Kaplan},
  {Cook}, {Van Sistine}, {Lau}, {Fremling}, {Gottlieb}, {Jencson}, {Adams},
  {Feindt}, {Hotokezaka}, {Ghosh}, {Perley}, {Yu}, {Piran}, {Allison},
  {Anupama}, {Balasubramanian}, {Bannister}, {Bally}, {Barnes}, {Barway},
  {Bellm}, {Bhalerao}, {Bhattacharya}, {Blagorodnova}, {Bloom}, {Brady},
  {Cannella}, {Chatterjee}, {Cenko}, {Cobb}, {Copperwheat}, {Corsi}, {De},
  {Dobie}, {Emery}, {Evans}, {Fox}, {Frail}, {Frohmaier}, {Goobar}, {Hallinan},
  {Harrison}, {Helou}, {Hinderer}, {Ho}, {Horesh}, {Ip}, {Itoh}, {Kasen},
  {Kim}, {Kuin}, {Kupfer}, {Lynch}, {Madsen}, {Mazzali}, {Miller}, {Mooley},
  {Murphy}, {Ngeow}, {Nichols}, {Nissanke}, {Nugent}, {Ofek}, {Qi}, {Quimby},
  {Rosswog}, {Rusu}, {Sadler}, {Schmidt}, {Sollerman}, {Steele}, {Williamson},
  {Xu}, {Yan}, {Yatsu}, {Zhang}, \& {Zhao}}]{GW170817mansi}
{Kasliwal}, M.~M., {Nakar}, E., {Singer}, L.~P., {et~al.} 2017, Science, 358,
  1559

\bibitem[{{Kennicutt} {et~al.}(1994){Kennicutt}, {Tamblyn}, \&
  {Congdon}}]{ken94}
{Kennicutt}, Jr., R.~C., {Tamblyn}, P., \& {Congdon}, C.~E. 1994, \apj, 435, 22

\bibitem[{{Klein} \& {Chevalier}(1978)}]{kle78}
{Klein}, R.~I., \& {Chevalier}, R.~A. 1978, \apjl, 223, L109

\bibitem[{{Komatsu} {et~al.}(2009){Komatsu}, {Dunkley}, {Nolta}, {Bennett},
  {Gold}, {Hinshaw}, {Jarosik}, {Larson}, {Limon}, {Page}, {Spergel},
  {Halpern}, {Hill}, {Kogut}, {Meyer}, {Tucker}, {Weiland}, {Wollack}, \&
  {Wright}}]{kom09}
{Komatsu}, E., {Dunkley}, J., {Nolta}, M.~R., {et~al.} 2009, \apjs, 180, 330

\bibitem[{{Kowalski} {et~al.}(2009){Kowalski}, {Hawley}, {Hilton}, {Becker},
  {West}, {Bochanski}, \& {Sesar}}]{kow09}
{Kowalski}, A.~F., {Hawley}, S.~L., {Hilton}, E.~J., {et~al.} 2009, \aj, 138,
  633

\bibitem[{{Law} {et~al.}(2009){Law}, {Kulkarni}, {Dekany}, {Ofek}, {Quimby},
  {Nugent}, {Surace}, {Grillmair}, {Bloom}, {Kasliwal}, {Bildsten}, {Brown},
  {Cenko}, {Ciardi}, {Croner}, {Djorgovski}, {van Eyken}, {Filippenko}, {Fox},
  {Gal-Yam}, {Hale}, {Hamam}, {Helou}, {Henning}, {Howell}, {Jacobsen},
  {Laher}, {Mattingly}, {McKenna}, {Pickles}, {Poznanski}, {Rahmer}, {Rau},
  {Rosing}, {Shara}, {Smith}, {Starr}, {Sullivan}, {Velur}, {Walters}, \&
  {Zolkower}}]{law09}
{Law}, N.~M., {Kulkarni}, S.~R., {Dekany}, R.~G., {et~al.} 2009, \pasp, 121,
  1395

\bibitem[{{Li} {et~al.}(2011){Li}, {Leaman}, {Chornock}, {Filippenko},
  {Poznanski}, {Ganeshalingam}, {Wang}, {Modjaz}, {Jha}, {Foley}, \&
  {Smith}}]{li11part2}
{Li}, W., {Leaman}, J., {Chornock}, R., {et~al.} 2011, \mnras, 412, 1441

\bibitem[{{Lien} {et~al.}(2014){Lien}, {Sakamoto}, {Gehrels}, {Palmer},
  {Barthelmy}, {Graziani}, \& {Cannizzo}}]{lie14}
{Lien}, A., {Sakamoto}, T., {Gehrels}, N., {et~al.} 2014, \apj, 783, 24

\bibitem[{{Mink}(2002)}]{min02}
{Mink}, D.~J. 2002, in Astronomical Society of the Pacific Conference Series,
  Vol. 281, Astronomical Data Analysis Software and Systems XI, ed. D.~A.
  {Bohlender}, D.~{Durand}, \& T.~H. {Handley}, 169

\bibitem[{{Miyazaki} {et~al.}(2002){Miyazaki}, {Komiyama}, {Sekiguchi},
  {Okamura}, {Doi}, {Furusawa}, {Hamabe}, {Imi}, {Kimura}, {Nakata}, {Okada},
  {Ouchi}, {Shimasaku}, {Yagi}, \& {Yasuda}}]{miy02}
{Miyazaki}, S., {Komiyama}, Y., {Sekiguchi}, M., {et~al.} 2002, \pasj, 54, 833

\bibitem[{{Miyazaki} {et~al.}(2006){Miyazaki}, {Komiyama}, {Nakaya}, {Doi},
  {Furusawa}, {Gillingham}, {Kamata}, {Takeshi}, \& {Nariai}}]{miy06}
{Miyazaki}, S., {Komiyama}, Y., {Nakaya}, H., {et~al.} 2006, in Presented at
  the Society of Photo-Optical Instrumentation Engineers (SPIE) Conference,
  Vol. 6269, Society of Photo-Optical Instrumentation Engineers (SPIE)
  Conference Series

\bibitem[{{Miyazaki} {et~al.}(2012){Miyazaki}, {Komiyama}, {Nakaya}, {Kamata},
  {Doi}, {Hamana}, {Karoji}, {Furusawa}, {Kawanomoto}, {Morokuma}, {Ishizuka},
  {Nariai}, {Tanaka}, {Uraguchi}, {Utsumi}, {Obuchi}, {Okura}, {Oguri},
  {Takata}, {Tomono}, {Kurakami}, {Namikawa}, {Usuda}, {Yamanoi}, {Terai},
  {Uekiyo}, {Yamada}, {Koike}, {Aihara}, {Fujimori}, {Mineo}, {Miyatake},
  {Yasuda}, {Nishizawa}, {Saito}, {Tanaka}, {Uchida}, {Katayama}, {Wang},
  {Chen}, {Lupton}, {Loomis}, {Bickerton}, {Price}, {Gunn}, {Suzuki},
  {Miyazaki}, {Muramatsu}, {Yamamoto}, {Endo}, {Ezaki}, {Itoh}, {Miwa},
  {Yokota}, {Matsuda}, {Ebinuma}, \& {Takeshi}}]{miy12}
{Miyazaki}, S., {Komiyama}, Y., {Nakaya}, H., {et~al.} 2012, in Society of
  Photo-Optical Instrumentation Engineers (SPIE) Conference Series, Vol. 8446,
  Society of Photo-Optical Instrumentation Engineers (SPIE) Conference Series,
  0

\bibitem[{{Moriya} {et~al.}(2018){Moriya}, {F{\"o}rster}, {Yoon},
  {Gr{\"a}fener}, \& {Blinnikov}}]{moriya18}
{Moriya}, T.~J., {F{\"o}rster}, F., {Yoon}, S.-C., {Gr{\"a}fener}, G., \&
  {Blinnikov}, S.~I. 2018, \mnras, 476, 2840

\bibitem[{{Morokuma} {et~al.}(2014){Morokuma}, {Tominaga}, {Tanaka}, {Mori},
  {Matsumoto}, {Kikuchi}, {Shibata}, {Sako}, {Aoki}, {Doi}, {Kobayashi},
  {Maehara}, {Matsunaga}, {Mito}, {Miyata}, {Nakada}, {Soyano}, {Tarusawa},
  {Miyazaki}, {Nakata}, {Okada}, {Sarugaku}, {Richmond}, {Akitaya}, {Aldering},
  {Arimatsu}, {Contreras}, {Horiuchi}, {Hsiao}, {Itoh}, {Iwata}, {Kawabata},
  {Kawai}, {Kitagawa}, {Kokubo}, {Kuroda}, {Mazzali}, {Misawa}, {Moritani},
  {Morrell}, {Okamoto}, {Pavlyuk}, {Phillips}, {Pian}, {Sahu}, {Saito}, {Sano},
  {Stritzinger}, {Tachibana}, {Taddia}, {Takaki}, {Tateuchi}, {Tomita},
  {Tsvetkov}, {Ui}, {Ukita}, {Urata}, {Walker}, \& {Yoshii}}]{mor14}
{Morokuma}, T., {Tominaga}, N., {Tanaka}, M., {et~al.} 2014, \pasj, 66, 114

\bibitem[{{Morozova} {et~al.}(2016){Morozova}, {Piro}, {Renzo}, \&
  {Ott}}]{mor16}
{Morozova}, V., {Piro}, A.~L., {Renzo}, M., \& {Ott}, C.~D. 2016, \apj, 829,
  109

\bibitem[{{Morozova} {et~al.}(2017){Morozova}, {Piro}, \& {Valenti}}]{mor17}
{Morozova}, V., {Piro}, A.~L., \& {Valenti}, S. 2017, \apj, 838, 28

\bibitem[{{Morrissey} {et~al.}(2005){Morrissey}, {Schiminovich}, {Barlow},
  {Martin}, {Blakkolb}, {Conrow}, {Cooke}, {Erickson}, {Fanson}, {Friedman},
  {Grange}, {Jelinsky}, {Lee}, {Liu}, {Mazer}, {McLean}, {Milliard}, {Randall},
  {Schmitigal}, {Sen}, {Siegmund}, {Surber}, {Vaughan}, {Viton}, {Welsh},
  {Bianchi}, {Byun}, {Donas}, {Forster}, {Heckman}, {Lee}, {Madore}, {Malina},
  {Neff}, {Rich}, {Small}, {Szalay}, \& {Wyder}}]{mor05}
{Morrissey}, P., {Schiminovich}, D., {Barlow}, T.~A., {et~al.} 2005, \apjl,
  619, L7

\bibitem[{{Morrissey} {et~al.}(2007){Morrissey}, {Conrow}, {Barlow}, {Small},
  {Seibert}, {Wyder}, {Budav{\'a}ri}, {Arnouts}, {Friedman}, {Forster},
  {Martin}, {Neff}, {Schiminovich}, {Bianchi}, {Donas}, {Heckman}, {Lee},
  {Madore}, {Milliard}, {Rich}, {Szalay}, {Welsh}, \& {Yi}}]{mor07}
{Morrissey}, P., {Conrow}, T., {Barlow}, T.~A., {et~al.} 2007, \apjs, 173, 682

\bibitem[{{Nagao} {et~al.}(2006){Nagao}, {Maiolino}, \& {Marconi}}]{nagao06}
{Nagao}, T., {Maiolino}, R., \& {Marconi}, A. 2006, \aap, 459, 85

\bibitem[{{Nakar} \& {Sari}(2010)}]{nak10}
{Nakar}, E., \& {Sari}, R. 2010, \apj, 725, 904

\bibitem[{{Nugent} {et~al.}(2011){Nugent}, {Sullivan}, {Cenko}, {Thomas},
  {Kasen}, {Howell}, {Bersier}, {Bloom}, {Kulkarni}, {Kandrashoff},
  {Filippenko}, {Silverman}, {Marcy}, {Howard}, {Isaacson}, {Maguire},
  {Suzuki}, {Tarlton}, {Pan}, {Bildsten}, {Fulton}, {Parrent}, {Sand},
  {Podsiadlowski}, {Bianco}, {Dilday}, {Graham}, {Lyman}, {James}, {Kasliwal},
  {Law}, {Quimby}, {Hook}, {Walker}, {Mazzali}, {Pian}, {Ofek}, {Gal-Yam}, \&
  {Poznanski}}]{nug11}
{Nugent}, P.~E., {Sullivan}, M., {Cenko}, S.~B., {et~al.} 2011, \nat, 480, 344

\bibitem[{{Ofek} {et~al.}(2010){Ofek}, {Rabinak}, {Neill}, {Arcavi}, {Cenko},
  {Waxman}, {Kulkarni}, {Gal-Yam}, {Nugent}, {Bildsten}, {Bloom}, {Filippenko},
  {Forster}, {Howell}, {Jacobsen}, {Kasliwal}, {Law}, {Martin}, {Poznanski},
  {Quimby}, {Shen}, {Sullivan}, {Dekany}, {Rahmer}, {Hale}, {Smith},
  {Zolkower}, {Velur}, {Walters}, {Henning}, {Bui}, \& {McKenna}}]{ofe10}
{Ofek}, E.~O., {Rabinak}, I., {Neill}, J.~D., {et~al.} 2010, \apj, 724, 1396

\bibitem[{{Pei}(1992)}]{pei92}
{Pei}, Y.~C. 1992, \apj, 395, 130

\bibitem[{{Perley} {et~al.}(2019){Perley}, {Mazzali}, {Yan}, {Cenko}, {Gezari},
  {Taggart}, {Blagorodnova}, {Fremling}, {Mockler}, {Singh}, {Tominaga},
  {Tanaka}, {Watson}, {Ahumada}, {Anupama}, {Ashall}, {Becerra}, {Bersier},
  {Bhalerao}, {Bloom}, {Butler}, {Copperwheat}, {Coughlin}, {De}, {Drake},
  {Duev}, {Frederick}, {Gonz{\'a}lez}, {Goobar}, {Heida}, {Ho}, {Horst},
  {Hung}, {Itoh}, {Jencson}, {Kasliwal}, {Kawai}, {Khanam}, {Kulkarni},
  {Kumar}, {Kumar}, {Kutyrev}, {Lee}, {Maeda}, {Mahabal}, {Murata}, {Neill},
  {Ngeow}, {Penprase}, {Pian}, {Quimby}, {Ramirez-Ruiz}, {Richer},
  {Rom{\'a}n-Z{\'u}{\~n}iga}, {Sahu}, {Srivastav}, {Socia}, {Sollerman},
  {Tachibana}, {Taddia}, {Tinyanont}, {Troja}, {Ward}, {Wee}, \& {Yu}}]{per19}
{Perley}, D.~A., {Mazzali}, P.~A., {Yan}, L., {et~al.} 2019, \mnras, 484, 1031

\bibitem[{{Pian} {et~al.}(2017){Pian}, {D'Avanzo}, {Benetti}, {Branchesi},
  {Brocato}, {Campana}, {Cappellaro}, {Covino}, {D'Elia}, {Fynbo}, {Getman},
  {Ghirlanda}, {Ghisellini}, {Grado}, {Greco}, {Hjorth}, {Kouveliotou},
  {Levan}, {Limatola}, {Malesani}, {Mazzali}, {Melandri}, {M{\o}ller},
  {Nicastro}, {Palazzi}, {Piranomonte}, {Rossi}, {Salafia}, {Selsing},
  {Stratta}, {Tanaka}, {Tanvir}, {Tomasella}, {Watson}, {Yang}, {Amati},
  {Antonelli}, {Ascenzi}, {Bernardini}, {Bo{\"e}r}, {Bufano}, {Bulgarelli},
  {Capaccioli}, {Casella}, {Castro-Tirado}, {Chassande-Mottin}, {Ciolfi},
  {Copperwheat}, {Dadina}, {De Cesare}, {di Paola}, {Fan}, {Gendre},
  {Giuffrida}, {Giunta}, {Hunt}, {Israel}, {Jin}, {Kasliwal}, {Klose}, {Lisi},
  {Longo}, {Maiorano}, {Mapelli}, {Masetti}, {Nava}, {Patricelli}, {Perley},
  {Pescalli}, {Piran}, {Possenti}, {Pulone}, {Razzano}, {Salvaterra},
  {Schipani}, {Spera}, {Stamerra}, {Stella}, {Tagliaferri}, {Testa}, {Troja},
  {Turatto}, {Vergani}, \& {Vergani}}]{GW170817pian}
{Pian}, E., {D'Avanzo}, P., {Benetti}, S., {et~al.} 2017, \nat, 551, 67

\bibitem[{{Pier} {et~al.}(2003){Pier}, {Munn}, {Hindsley}, {Hennessy}, {Kent},
  {Lupton}, \& {Ivezi{\'c}}}]{pie03}
{Pier}, J.~R., {Munn}, J.~A., {Hindsley}, R.~B., {et~al.} 2003, \aj, 125, 1559

\bibitem[{{Prentice} {et~al.}(2018){Prentice}, {Maguire}, {Smartt}, {Magee},
  {Schady}, {Sim}, {Chen}, {Clark}, {Colin}, {Fulton}, {McBrien}, {O'Neill},
  {Smith}, {Ashall}, {Chambers}, {Denneau}, {Flewelling}, {Heinze}, {Holoien},
  {Huber}, {Kochanek}, {Mazzali}, {Prieto}, {Rest}, {Shappee}, {Stalder},
  {Stanek}, {Stritzinger}, {Thompson}, \& {Tonry}}]{pre18}
{Prentice}, S.~J., {Maguire}, K., {Smartt}, S.~J., {et~al.} 2018, \apjl, 865,
  L3

\bibitem[{{Pritchard} {et~al.}(2014){Pritchard}, {Roming}, {Brown}, {Bayless},
  \& {Frey}}]{pri14}
{Pritchard}, T.~A., {Roming}, P.~W.~A., {Brown}, P.~J., {Bayless}, A.~J., \&
  {Frey}, L.~H. 2014, \apj, 787, 157

\bibitem[{{Rau} {et~al.}(2009){Rau}, {Kulkarni}, {Law}, {Bloom}, {Ciardi},
  {Djorgovski}, {Fox}, {Gal-Yam}, {Grillmair}, {Kasliwal}, {Nugent}, {Ofek},
  {Quimby}, {Reach}, {Shara}, {Bildsten}, {Cenko}, {Drake}, {Filippenko},
  {Helfand}, {Helou}, {Howell}, {Poznanski}, \& {Sullivan}}]{rau09}
{Rau}, A., {Kulkarni}, S.~R., {Law}, N.~M., {et~al.} 2009, \pasp, 121, 1334

\bibitem[{{Richardson} {et~al.}(2014){Richardson}, {Jenkins}, {Wright}, \&
  {Maddox}}]{ric14}
{Richardson}, D., {Jenkins}, III, R.~L., {Wright}, J., \& {Maddox}, L. 2014,
  \aj, 147, 118

\bibitem[{{Richmond} {et~al.}(1994){Richmond}, {Treffers}, {Filippenko},
  {Paik}, {Leibundgut}, {Schulman}, \& {Cox}}]{ric94}
{Richmond}, M.~W., {Treffers}, R.~R., {Filippenko}, A.~V., {et~al.} 1994, \aj,
  107, 1022

\bibitem[{{Richmond} {et~al.}(1996){Richmond}, {van Dyk}, {Ho}, {Peng}, {Paik},
  {Treffers}, {Filippenko}, {Bustamante-Donas}, {Moeller}, {Pawellek},
  {Tartara}, \& {Spence}}]{ric96}
{Richmond}, M.~W., {van Dyk}, S.~D., {Ho}, W., {et~al.} 1996, \aj, 111, 327

\bibitem[{{Rubin} \& {Gal-Yam}(2017)}]{rub17}
{Rubin}, A., \& {Gal-Yam}, A. 2017, \apj, 848, 8

\bibitem[{{Sari} {et~al.}(1998){Sari}, {Piran}, \& {Narayan}}]{sar98}
{Sari}, R., {Piran}, T., \& {Narayan}, R. 1998, \apjl, 497, L17

\bibitem[{{Schawinski} {et~al.}(2008){Schawinski}, {Justham}, {Wolf},
  {Podsiadlowski}, {Sullivan}, {Steenbrugge}, {Bell}, {R{\"o}ser}, {Walker},
  {Astier}, {Balam}, {Balland}, {Carlberg}, {Conley}, {Fouchez}, {Guy},
  {Hardin}, {Hook}, {Howell}, {Pain}, {Perrett}, {Pritchet}, {Regnault}, \&
  {Yi}}]{sch08}
{Schawinski}, K., {Justham}, S., {Wolf}, C., {et~al.} 2008, Science, 321, 223

\bibitem[{{Schlafly} \& {Finkbeiner}(2011)}]{sch11}
{Schlafly}, E.~F., \& {Finkbeiner}, D.~P. 2011, \apj, 737, 103

\bibitem[{{Schmidt}(1968)}]{sch68}
{Schmidt}, M. 1968, \apj, 151, 393

\bibitem[{{Scolnic} {et~al.}(2018){Scolnic}, {Kessler}, {Brout},
  {Cowperthwaite}, {Soares-Santos}, {Annis}, {Herner}, {Chen}, {Sako},
  {Doctor}, {Butler}, {Palmese}, {Diehl}, {Frieman}, {Holz}, {Berger},
  {Chornock}, {Villar}, {Nicholl}, {Biswas}, {Hounsell}, {Foley}, {Metzger},
  {Rest}, {Garc{\'{\i}}a-Bellido}, {M{\"o}ller}, {Nugent}, {Abbott}, {Abdalla},
  {Allam}, {Bechtol}, {Benoit-L{\'e}vy}, {Bertin}, {Brooks}, {Buckley-Geer},
  {Carnero Rosell}, {Carrasco Kind}, {Carretero}, {Castander}, {Cunha},
  {D'Andrea}, {da Costa}, {Davis}, {Doel}, {Drlica-Wagner}, {Eifler},
  {Flaugher}, {Fosalba}, {Gaztanaga}, {Gerdes}, {Gruen}, {Gruendl}, {Gschwend},
  {Gutierrez}, {Hartley}, {Honscheid}, {James}, {Johnson}, {Johnson}, {Krause},
  {Kuehn}, {Kuhlmann}, {Lahav}, {Li}, {Lima}, {Maia}, {March}, {Marshall},
  {Menanteau}, {Miquel}, {Neilsen}, {Plazas}, {Sanchez}, {Scarpine},
  {Schubnell}, {Sevilla-Noarbe}, {Smith}, {Smith}, {Sobreira}, {Suchyta},
  {Swanson}, {Tarle}, {Thomas}, {Tucker}, {Walker}, \& {DES
  Collaboration}}]{sco18}
{Scolnic}, D., {Kessler}, R., {Brout}, D., {et~al.} 2018, \apjl, 852, L3

\bibitem[{{Shappee} {et~al.}(2017){Shappee}, {Simon}, {Drout}, {Piro},
  {Morrell}, {Prieto}, {Kasen}, {Holoien}, {Kollmeier}, {Kelson}, {Coulter},
  {Foley}, {Kilpatrick}, {Siebert}, {Madore}, {Murguia-Berthier}, {Pan},
  {Prochaska}, {Ramirez-Ruiz}, {Rest}, {Adams}, {Alatalo}, {Ba{\~n}ados},
  {Baughman}, {Bernstein}, {Bitsakis}, {Boutsia}, {Bravo}, {Di Mille}, {Higgs},
  {Ji}, {Maravelias}, {Marshall}, {Placco}, {Prieto}, \&
  {Wan}}]{GW170817shappee}
{Shappee}, B.~J., {Simon}, J.~D., {Drout}, M.~R., {et~al.} 2017, Science, 358,
  1574

\bibitem[{{Smartt}(2009)}]{sma09b}
{Smartt}, S.~J. 2009, \araa, 47, 63

\bibitem[{{Smartt} {et~al.}(2017){Smartt}, {Chen}, {Jerkstrand}, {Coughlin},
  {Kankare}, {Sim}, {Fraser}, {Inserra}, {Maguire}, {Chambers}, {Huber},
  {Kr{\"u}hler}, {Leloudas}, {Magee}, {Shingles}, {Smith}, {Young}, {Tonry},
  {Kotak}, {Gal-Yam}, {Lyman}, {Homan}, {Agliozzo}, {Anderson}, {Angus},
  {Ashall}, {Barbarino}, {Bauer}, {Berton}, {Botticella}, {Bulla}, {Bulger},
  {Cannizzaro}, {Cano}, {Cartier}, {Cikota}, {Clark}, {De Cia}, {Della Valle},
  {Denneau}, {Dennefeld}, {Dessart}, {Dimitriadis}, {Elias-Rosa}, {Firth},
  {Flewelling}, {Fl{\"o}rs}, {Franckowiak}, {Frohmaier}, {Galbany},
  {Gonz{\'a}lez-Gait{\'a}n}, {Greiner}, {Gromadzki}, {Guelbenzu},
  {Guti{\'e}rrez}, {Hamanowicz}, {Hanlon}, {Harmanen}, {Heintz}, {Heinze},
  {Hernandez}, {Hodgkin}, {Hook}, {Izzo}, {James}, {Jonker}, {Kerzendorf},
  {Klose}, {Kostrzewa-Rutkowska}, {Kowalski}, {Kromer}, {Kuncarayakti},
  {Lawrence}, {Lowe}, {Magnier}, {Manulis}, {Martin-Carrillo}, {Mattila},
  {McBrien}, {M{\"u}ller}, {Nordin}, {O'Neill}, {Onori}, {Palmerio},
  {Pastorello}, {Patat}, {Pignata}, {Podsiadlowski}, {Pumo}, {Prentice}, {Rau},
  {Razza}, {Rest}, {Reynolds}, {Roy}, {Ruiter}, {Rybicki}, {Salmon}, {Schady},
  {Schultz}, {Schweyer}, {Seitenzahl}, {Smith}, {Sollerman}, {Stalder},
  {Stubbs}, {Sullivan}, {Szegedi}, {Taddia}, {Taubenberger}, {Terreran}, {van
  Soelen}, {Vos}, {Wainscoat}, {Walton}, {Waters}, {Weiland}, {Willman},
  {Wiseman}, {Wright}, {Wyrzykowski}, \& {Yaron}}]{GW170817smartt}
{Smartt}, S.~J., {Chen}, T.-W., {Jerkstrand}, A., {et~al.} 2017, \nat, 551, 75

\bibitem[{{Svensson} {et~al.}(2010){Svensson}, {Levan}, {Tanvir}, {Fruchter},
  \& {Strolger}}]{sve10}
{Svensson}, K.~M., {Levan}, A.~J., {Tanvir}, N.~R., {Fruchter}, A.~S., \&
  {Strolger}, L.-G. 2010, \mnras, 405, 57

\bibitem[{{Tanaka}(2015)}]{tanaka15photoz}
{Tanaka}, M. 2015, \apj, 801, 20

\bibitem[{{Tanaka} {et~al.}(2016){Tanaka}, {Tominaga}, {Morokuma}, {Yasuda},
  {Furusawa}, {Baklanov}, {Blinnikov}, {Moriya}, {Doi}, {Jiang}, {Kato},
  {Kikuchi}, {Kuncarayakti}, {Nagao}, {Nomoto}, \& {Taniguchi}}]{tanaka16}
{Tanaka}, M., {Tominaga}, N., {Morokuma}, T., {et~al.} 2016, \apj, 819, 5

\bibitem[{{Tody}(1986)}]{tod86}
{Tody}, D. 1986, in Society of Photo-Optical Instrumentation Engineers (SPIE)
  Conference Series, Vol. 627, \procspie, ed. D.~L. {Crawford}, 733

\bibitem[{{Tody}(1993)}]{tod93}
{Tody}, D. 1993, in Astronomical Society of the Pacific Conference Series,
  Vol.~52, Astronomical Data Analysis Software and Systems II, ed. R.~J.
  {Hanisch}, R.~J.~V. {Brissenden}, \& J.~{Barnes}, 173

\bibitem[{{Tolstov} {et~al.}(2013){Tolstov}, {Blinnikov}, \&
  {Nadyozhin}}]{tol13}
{Tolstov}, A.~G., {Blinnikov}, S.~I., \& {Nadyozhin}, D.~K. 2013, \mnras, 429,
  3181

\bibitem[{{Tominaga} {et~al.}(2009){Tominaga}, {Blinnikov}, {Baklanov},
  {Morokuma}, {Nomoto}, \& {Suzuki}}]{tom09b}
{Tominaga}, N., {Blinnikov}, S., {Baklanov}, P., {et~al.} 2009, \apjl, 705, L10

\bibitem[{{Tominaga} {et~al.}(2011){Tominaga}, {Morokuma}, {Blinnikov},
  {Baklanov}, {Sorokina}, \& {Nomoto}}]{tom11}
{Tominaga}, N., {Morokuma}, T., {Blinnikov}, S.~I., {et~al.} 2011, \apjs, 193,
  20

\bibitem[{{Tominaga} {et~al.}(2014{\natexlab{a}}){Tominaga}, {Morokuma},
  {Tanaka}, {Yasuda}, {Furusawa}, {Jiang}, {Miyazaki}, {Moriya}, {Noumaru},
  {Schubert}, \& {Takata}}]{tom14atel}
{Tominaga}, N., {Morokuma}, T., {Tanaka}, M., {et~al.} 2014{\natexlab{a}}, The
  Astronomer's Telegram, 6291, 1

\bibitem[{{Tominaga} {et~al.}(2014{\natexlab{b}}){Tominaga}, {Morokuma},
  {Tanaka}, {Yasuda}, {Furusawa}, {Jiang}, {Okabe}, {Futamase}, {Miyazaki},
  {Moriya}, {Noumaru}, {Schubert}, \& {Takata}}]{tom14atel2}
---. 2014{\natexlab{b}}, The Astronomer's Telegram, 6763, 1

\bibitem[{{Tominaga} {et~al.}(2015{\natexlab{a}}){Tominaga}, {Morokuma},
  {Tanaka}, {Jiang}, {Kato}, {Taniguchi}, {Yasuda}, {Furusawa}, {Okabe},
  {Futamase}, {Miyazaki}, {Moriya}, {Noumaru}, {Schubert}, \&
  {Takata}}]{tom15atel}
---. 2015{\natexlab{a}}, The Astronomer's Telegram, 7565, 1

\bibitem[{{Tominaga} {et~al.}(2015{\natexlab{b}}){Tominaga}, {Morokuma},
  {Tanaka}, {Jiang}, {Kato}, {Taniguchi}, {Yasuda}, {Furusawa}, {Okabe},
  {Futamase}, {Miyazaki}, {Moriya}, {Noumaru}, {Schubert}, \&
  {Takata}}]{tom15atel2}
---. 2015{\natexlab{b}}, The Astronomer's Telegram, 7565, 1

\bibitem[{{Tominaga} {et~al.}(2018{\natexlab{a}}){Tominaga}, {Niino}, {Totani},
  {Yasuda}, {Furusawa}, {Tanaka}, {Bhandari}, {Dodson}, {Keane}, {Morokuma},
  {Petroff}, \& {Possenti}}]{tom18frb}
{Tominaga}, N., {Niino}, Y., {Totani}, T., {et~al.} 2018{\natexlab{a}}, \pasj,
  70, 103

\bibitem[{{Tominaga} {et~al.}(2018{\natexlab{b}}){Tominaga}, {Tanaka},
  {Morokuma}, {Utsumi}, {Yamaguchi}, {Yasuda}, {Tanaka}, {Yoshida},
  {Fujiyoshi}, {Furusawa}, {Kawabata}, {Lee}, {Motohara}, {Ohsawa}, {Ohta},
  {Terai}, {Abe}, {Aoki}, {Asakura}, {Barway}, {Bond}, {Fujisawa}, {Honda},
  {Ioka}, {Itoh}, {Kawai}, {Kim}, {Koshimoto}, {Matsubayashi}, {Miyazaki},
  {Saito}, {Sekiguchi}, {Sumi}, \& {Tristram}}]{tom18gw170817}
{Tominaga}, N., {Tanaka}, M., {Morokuma}, T., {et~al.} 2018{\natexlab{b}},
  \pasj, 70, 28

\bibitem[{{Totani} \& {Panaitescu}(2002)}]{tot02}
{Totani}, T., \& {Panaitescu}, A. 2002, \apj, 576, 120

\bibitem[{{Tresse} {et~al.}(2002){Tresse}, {Maddox}, {Le F{\`e}vre}, \&
  {Cuby}}]{tre02}
{Tresse}, L., {Maddox}, S.~J., {Le F{\`e}vre}, O., \& {Cuby}, J.-G. 2002,
  \mnras, 337, 369

\bibitem[{{Tsvetkov} {et~al.}(2012){Tsvetkov}, {Volkov}, {Sorokina},
  {Blinnikov}, {Pavlyuk}, \& {Borisov}}]{tsv12}
{Tsvetkov}, D.~Y., {Volkov}, I.~M., {Sorokina}, E., {et~al.} 2012, Peremennye
  Zvezdy, 32, arXiv:1207.2241

\bibitem[{{Utsumi} {et~al.}(2017){Utsumi}, {Tanaka}, {Tominaga}, {Yoshida},
  {Barway}, {Nagayama}, {Zenko}, {Aoki}, {Fujiyoshi}, {Furusawa}, {Kawabata},
  {Koshida}, {Lee}, {Morokuma}, {Motohara}, {Nakata}, {Ohsawa}, {Ohta},
  {Okita}, {Tajitsu}, {Tanaka}, {Terai}, {Yasuda}, {Abe}, {Asakura}, {Bond},
  {Miyazaki}, {Sumi}, {Tristram}, {Honda}, {Itoh}, {Itoh}, {Kawabata},
  {Morihana}, {Nagashima}, {Nakaoka}, {Ohshima}, {Takahashi}, {Takayama},
  {Aoki}, {Baar}, {Doi}, {Finet}, {Kanda}, {Kawai}, {Kim}, {Kuroda}, {Liu},
  {Matsubayashi}, {Murata}, {Nagai}, {Saito}, {Saito}, {Sako}, {Sekiguchi},
  {Tamura}, {Tanaka}, {Uemura}, \& {Yamaguchi}}]{utsumi17}
{Utsumi}, Y., {Tanaka}, M., {Tominaga}, N., {et~al.} 2017, \pasj, 69, 101

\bibitem[{{Utsumi} {et~al.}(2018){Utsumi}, {Tominaga}, {Tanaka}, {Morokuma},
  {Yoshida}, {Asakura}, {Finet}, {Furusawa}, {Kawabata}, {Liu}, {Matsubayashi},
  {Moritani}, {Motohara}, {Nakata}, {Ohta}, {Terai}, {Uemura}, \&
  {Yasuda}}]{uts18gw151226}
{Utsumi}, Y., {Tominaga}, N., {Tanaka}, M., {et~al.} 2018, \pasj, 70, 1

\bibitem[{{Valenti} {et~al.}(2017){Valenti}, {David}, {Sand}, {Yang},
  {Cappellaro}, {Tartaglia}, {Corsi}, {Jha}, {Reichart}, {Haislip}, \&
  {Kouprianov}}]{GW170817valenti}
{Valenti}, S., {David}, {Sand}, J., {et~al.} 2017, \apjl, 848, L24

\bibitem[{{Villar} {et~al.}(2017){Villar}, {Guillochon}, {Berger}, {Metzger},
  {Cowperthwaite}, {Nicholl}, {Alexander}, {Blanchard}, {Chornock},
  {Eftekhari}, {Fong}, {Margutti}, \& {Williams}}]{GW170817OptSum}
{Villar}, V.~A., {Guillochon}, J., {Berger}, E., {et~al.} 2017, \apjl, 851, L21

\bibitem[{{Waxman} {et~al.}(2007){Waxman}, {M{\'e}sz{\'a}ros}, \&
  {Campana}}]{wax07}
{Waxman}, E., {M{\'e}sz{\'a}ros}, P., \& {Campana}, S. 2007, \apj, 667, 351

\bibitem[{{Yaron} {et~al.}(2017){Yaron}, {Perley}, {Gal-Yam}, {Groh}, {Horesh},
  {Ofek}, {Kulkarni}, {Sollerman}, {Fransson}, {Rubin}, {Szabo}, {Sapir},
  {Taddia}, {Cenko}, {Valenti}, {Arcavi}, {Howell}, {Kasliwal}, {Vreeswijk},
  {Khazov}, {Fox}, {Cao}, {Gnat}, {Kelly}, {Nugent}, {Filippenko}, {Laher},
  {Wozniak}, {Lee}, {Rebbapragada}, {Maguire}, {Sullivan}, \&
  {Soumagnac}}]{yar17}
{Yaron}, O., {Perley}, D.~A., {Gal-Yam}, A., {et~al.} 2017, Nature Physics, 13,
  510

\end{thebibliography}

\end{document}